\documentclass[smallextended,12pt,psfig,a4]{amsart}

\usepackage[latin1]{inputenc}
\usepackage{graphicx,amsmath,dsfont}
\usepackage{indentfirst}
\usepackage{amssymb}
\usepackage{amsfonts}
\usepackage{graphicx}
\usepackage{array}
\usepackage{amsthm}
\usepackage{float}
\usepackage{multirow}
\usepackage[normalem]{ulem}
\usepackage{braket}
\usepackage{lastpage}
\usepackage{subcaption}
\usepackage{indentfirst}
\usepackage{microtype}
\usepackage{fancyhdr}
\usepackage[all]{xy}
\usepackage{blkarray, bigstrut}
\usepackage[toc,page]{appendix}
\usepackage{amsmath}               
\usepackage{amsfonts}              
\usepackage{amsthm}                
\usepackage{esint}
\usepackage{graphics}
\usepackage{verbatim}
\usepackage{tikz}

\usepackage[textwidth = 155mm]{geometry}
\usetikzlibrary{automata, arrows}

\def\bma{\begin{bmatrix}}
\def\ema{\end{bmatrix}}
\def\bex{\begin{example}}
\def\eex{\end{example}}
\def\beq{\begin{equation}}
\def\eeq{\end{equation}}

\def\qee{\begin{flushright} $\Diamond$ \end{flushright}}

\theoremstyle{plain}
\newtheorem{teo}{Theorem}[section]
\newtheorem{lema}[teo]{Lemma}

\newtheorem{defi}[teo]{Definition}
\newtheorem{prop}[teo]{Proposition}

\theoremstyle{definition}
\newtheorem{remark}{Remark}
\newtheorem{ex}[teo]{Example}
\newtheorem{REM}[teo]{Remark}

\begin{document}
\pagestyle{plain}

\title{Site Recurrence for continuous-time open quantum walks on the line}

\maketitle

\begin{center}\author{Newton Loebens\footnote{Email: newtonloebens@gmail.com}}\end{center}

\centerline{\footnotesize\it
Instituto de Matem\'atica e Estat\'istica, Universidade Federal do Rio Grande do Sul.}
\centerline{\footnotesize\it
Porto Alegre, RS  91509-900 Brazil.}
\baselineskip=10pt
\vspace*{10pt}
\vspace*{0.225truein}

\vspace*{0.21truein}

\begin{abstract}
In recent years, several properties and recurrence criteria of discrete-time open quantum walks (OQWs) have been presented. Recently, Pellegrini introduced continuous-time open quantum walks (CTOQWs) as continuous-time natural limits of discrete-time OQWs. In this work, we study semifinite CTOQWs and some of their basic properties concerning statistics, such as transition probabilities and site recurrence. The notion of SJK-recurrence for CTOQWs is introduced, and it is shown to be equivalent to the traditional concept of recurrence. This statistic arises from the definition of $\delta$-skeleton of CTOQWs, which is a dynamic that allows us to obtain a discrete-time OQW in terms of a CTOQW. We present a complete criterion for site recurrence in the case of CTOQW induced by a coin of finite dimension with a set of vertices $\mathbb{Z}$ such that its auxiliary Lindblad operator has a single stationary state. Finally, we present a similar criterion that completes the case in which the internal degree of freedom of each site is of dimension 2.
\\
\\
\textit{Keywords}:
Continuous-time open quantum walks.
Lindblad generator.
Site recurrence.
SJK-recurrence.
Stationary state.
\end{abstract}

\section{Introduction}

The discrete-time model of open quantum walk (OQW) was presented by Attal et al. (\cite{attal}) by the implementation of appropriate completely positive (CP) maps, considering the theory of open quantum systems. As in the classic case, we can define the concept of recurrence of a vertex from the average number of returns to the vertex. However, unlike the classical case, the quantum version allows a vertex to be recurrent with respect to a certain initial density and transient (non-recurrent) with respect to another initial density, thus any recurrence criteria must be more sophisticated.

Regarding the recurrence types of OQWs, many works can be found in the literature describing recurrence properties (see \cite{CGL,dILL,glv}). Recently, Jacq and Lardizabal (\cite{JL2}) exhibited complete recurrence criteria for two classes of OQWs on the real-line induced by a coin, one for irreducible OQWs (see \cite{carbone1}) based on an auxiliary CP map for finite coins and the other for the case where the coin is two-dimensional. The recurrence type considered in their work concerns the mean number of returns to site $\ket{0},$ and therefore to any site since they analyzed the homogeneous OQW.

In this work, we present some basic statistical properties of continuous-time Markov chains (CTOQWs) and recurrence criteria for some classes of those random walks. Those CTOQWs were presented by Pellegrini (\cite{pele}) as dynamics arising as continuous-time limits of time-discrete OQWs. CTOQW displays a graph's continuous-time evolution in which a particle jumps from vertex to vertex at random intervals. Jump intensity is determined by the internal degrees of freedom, which are affected by the jump but continue to evolve between jumps. A quantum mechanical model can be used to justify the form of the intensity as well as the evolution of the internal degrees of freedom at jump times and between them in both circumstances.

Since the behavior of the position of a CTOQW is not Markovian, and consequently does not retain memory, some characteristics of the transition probabilities differ from those of classical random walks. The Chapman-Kolmogorov identity, for example, is no longer valid, although a generalization for the quantum version can be produced. The initial density operator carries the dynamic's memory, and all potential generalizations rely on an adjustment to the density operator. Remark that the joint distribution of the position at instant $t$ with the density operator at time $t$ is a Markov process, however, the position alone is not.

We will only be dealing with semifinite CTOQWS, and once the transition probabilities for this class of walks are clearly understood, we will discuss some basic transition probabilities for those CTOQWs. We also provide an equivalency result between the recurrence of CTOQWs and discrete-time OQWs by a discretization of the CTOQW.

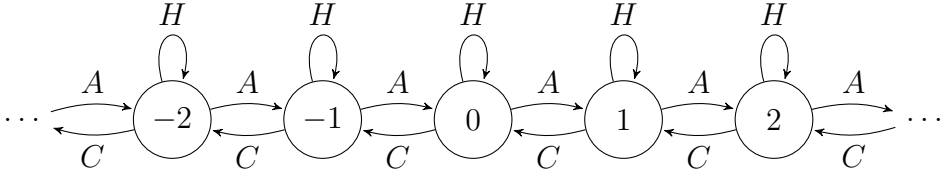
\begin{figure}[h!]
$$\begin{tikzpicture}[>=stealth',shorten >=1pt,auto,node distance=2cm]
\node[state] (q-2)      {$-2$};
\node[state]         (q-1) [right of=q-2]  {$-1$};
\node[state]         (q0) [right of=q-1]  {$0$};
\node[state]         (q1) [right of=q0]  {$1$};
\node[state]         (q2) [right of=q1]  {$2$};
\node         (qr) [right of=q2]  {$\ldots$};
\node         (ql) [left of=q-2]  {$\ldots$};
\path[->]          (ql)  edge         [bend left=15]  node[auto] {$A$}     (q-2);
\path[->]          (q-2)  edge         [bend left=15]   node[auto] {$A$}     (q-1);
\path[->]          (q-1)  edge         [bend left=15]   node[auto] {$A$}     (q0);
\path[->]          (q0)  edge         [bend left=15]   node[auto] {$A$}     (q1);
\path[->]          (q1)  edge         [bend left=15]   node[auto] {$A$}     (q2);
\path[->]          (q2)  edge         [bend left=15]   node[auto] {$A$}     (qr);
\path[->]          (qr)  edge         [bend left=15] node[auto] {$C$}       (q2);
\path[->]          (q2)  edge         [bend left=15]   node[auto] {$C$}     (q1);
\path[->]          (q1)  edge         [bend left=15]  node[auto] {$C$}      (q0);
\path[->]          (q0)  edge         [bend left=15]   node[auto] {$C$}     (q-1);
\path[->]          (q-1)  edge         [bend left=15]  node[auto] {$C$}      (q-2);
\path[->]          (q-2)  edge         [bend left=15]  node[auto] {$C$}      (ql);
\draw [->]       (q1) to[loop above]  node[auto] {$H$} (q1);
\draw [->]       (q-2) to[loop above]  node[auto] {$H$} (q-2);
\draw [->]       (q0) to[loop above]  node[auto] {$H$} (q0);
\draw [->]       (q-1) to[loop above]  node[auto] {$H$} (q-1);
\draw [->]       (q2) to[loop above]  node[auto] {$H$} (q2);
\end{tikzpicture}$$
\caption{Nearest-neighbor homogeneous walk on $\mathbb{Z}$.}
\label{graphZ}
\end{figure}

Concerning recurrence, we treat the one-dimensional case for nearest-neighbor homogeneous CTOQWs on the line, that is, with a set of vertices $\mathbb{Z}$, thus we will be considering the Lindblad generator introduced in \cite{bri} with $d=1$ and the rates of transition can be identified by the graph represented in Figure \ref{graphZ}, where $C,A,H$ are operators acting in the same finite Hilbert space and thus each site has the same internal degrees of freedom.

The operators $C$ and $A$ can be understood as rates of jumping to left and right, taking data from some vertex $\ket{i}$ and producing an output in $\ket{i-1}$ and $\ket{i+1},$ respectively, while $H$ represents a mixing of rates at $\ket{i}$ since it takes any data on vertex $\ket{i}$ and produces an output on $\ket{i},$ for each $i\in\mathbb{Z}$. Therefore, $H$ can contribute with the statistics of the walk, however, the walker must, at each step, jump to one of the two nearest-neighbors, and cannot jump in a loop to itself.

With this context well described, we intend to obtain a recurrence criterion for this kind of CTOQW in terms of the properties of the components of the Lindbladian: $C,\;A$ and $H,$ where $H$ is the Hamiltonian operator and a new operator to be considered to discuss recurrence when we compare it with the discrete-time model. As will be seen, some connections can be done with the discrete-time model, and thus we can apply some similar ideas to the continuous-time version. The great differences here are that we also use an auxiliary map, however, this map is no more a quantum channel, but a Lindblad operator acting in a finite-dimensional Hilbert space, and the statistics of the walk do not depend only on a dissipative part, but also on a Hamiltonian part of the Lindblad generator.

When the Hamiltonian operator $H$ is null, the evolution of the CTOQW depends only on the transition rates established by $C$ and $A$. On the other, a non-null $H$ will interfere with the stationary state of the auxiliary Lindblad operator that will be defined later and thus the recurrence may not depend exclusively on the conjugation generated by $C$ and $A$.

This article is organized as follows. In Section 2 we review CTOQWs, recurrence, and transition probability properties. We also present a new concept of recurrence based on the $\delta$-skeleton and show that this definition for site recurrence is equivalent to the standard definition. Section 3 is devoted to presenting recurrence criteria for CTOQWs induced by a finite coin such that the auxiliary Lindblad operator has a unique stationary state. In Section 4 we focus on the case where the coin has dimension $2,$ and we give a complete recurrence criterion for this class of CTOQWs. Section 5 illustrates the results with examples. Section 6 contains some of the most technic proofs of the paper.

\section{Continuous-Time Open Quantum Walks}

Let $\mathcal{H}$ denote a complex, separable Hilbert space with inner product $\langle\,\cdot\,|\,\cdot\,\rangle$, whose closed subspaces will be referred to as subspaces for short. The Banach algebra $\mathcal{B}(\mathcal{H})$ of bounded linear operators on $\mathcal{H}$ is the topological dual of its ideal $\mathcal{I}(\mathcal{H})$ of trace-class operators with trace norm
$$
\|\rho\|_1=\operatorname{\mathrm{Tr}}(|\rho|),
\qquad
|\rho|=\sqrt{\rho^*\rho},
$$
through the duality \cite[Lec. 6]{attal_lec}
\beq \label{eq:dual}
\langle \rho,X \rangle = \operatorname{\mathrm{Tr}}(\rho X),
\qquad
\rho\in\mathcal{I}(\mathcal{H}),
\qquad
X\in\mathcal{B}(\mathcal{H}),
\eeq
where the superscript ${}^*$ denotes the familiar adjoint operator.

If $\dim\mathcal{H}=k<\infty$, then $\mathcal{B}(\mathcal{H})=\mathcal{I}(\mathcal{H})$ is identified with the set of square matrices of order $k$, represented by $M_k(\mathbb{C}),$ and its identity operator will be denoted by $I_k.$ The duality \eqref{eq:dual} yields a useful characterization of the positivity of an operator $\rho\in\mathcal{I}(\mathcal{H})$,
\begin{equation*}\label{eq:pos-dual}
\rho\in\mathcal{I}(\mathcal{H}):
\quad
\rho\ge0 \; \Leftrightarrow \; \operatorname{\mathrm{Tr}}(\rho X)\ge0,
\quad
\forall X\in\mathcal{B}(\mathcal{H}),
\quad
X\ge0,
\end{equation*}
and similarly for the positivity of $X\in\mathcal{B}(\mathcal{H})$.

Moving toward the definition of CTOQWs, we recall a few basic results of one-parameter semigroups and completely positive maps. In particular, we discuss completely positive (CP) trace-preserving (TP) semigroups, which have a special generator operator. Those semigroups, called Continuous-time Open Quantum Walks (CTOQW's), are quantum generalizations of classical Markov continuous-time random walks.

Unlike the classical model, CTOQW describes the behavior of a random walk that retains some amount of memory, and this memory is encoded by a quantum state, which is a density operator acting on the associated Hilbert space.

\subsection{Basic Properties of Continuous-Time Open Quantum Walks}
An operator \textbf{semigroup}\index{semigroup} $\mathcal{T}$ on a Hilbert space $\mathcal{H}$ is a family of bounded linear operators $(T_t)$ acting
on $\mathcal{H}$, $t\geq 0,$ such that
\begin{equation}\label{def-semi}\nonumber
T_tT_s=T_{t+s},\;\forall s,t\in\mathbb{R}^+\;\mbox{ and }\;T_0=I_\mathcal{H}.
\end{equation}

If $t\mapsto T_t$ is continuous for the operator norm of $\mathcal{H}$, then $\mathcal{T}$ is said to be \textbf{uniformly
continuous}.\index{uniformly continuous semigroup} This class of semigroups is characterized by the following result:

\begin{teo}[\cite{bratteli}, page 161]
The following assertions are equivalent for a semigroup $\mathcal{T}$ on $\mathcal{H}:$
\begin{enumerate}
  \item $\mathcal{T}$ is uniformly continuous;
  \item There exists a bounded linear operator $L$ on $\mathcal{H}$ such that
  \begin{equation}\label{defi-ger}\nonumber
  T_t=e^{tL},\;\forall t\geq 0.
  \end{equation}
Further, if the equivalent conditions (1) and (2) are satisfied, then
\begin{equation}\label{defi-gerlim}\nonumber
L=\lim_{t\rightarrow\infty}\frac{1}{t}(T_t-I_\mathcal{H}).
\end{equation}
\end{enumerate}

The operator $L$ is called the \textbf{generator} of $\mathcal{T}.$ \index{generator of $\mathcal{T}$}
\end{teo}

In our subsequent discussion, we consider a countable set $V$ and a Hilbert space $\mathcal{H}$ of the form
\begin{equation}\label{H}
\mathcal{H}=\bigoplus_{i\in V}\mathfrak{h}_i,
\end{equation}
where each $\mathfrak{h}_i$ represents a separable Hilbert space.

A semigroup $\mathcal{T}:=(\mathcal{T}_t)_{t\geq 0}$ of CP-TP maps acting on the set of trace-class operators on $\mathcal{H},$ denoted $\mathcal{I}_1(\mathcal{H}),$ is called a \textbf{Quantum Markov Semigroup} (QMS)\index{Quantum Markov Semigroup (QMS)} on  $\mathcal{I}_1(\mathcal{H}).$ When $\lim_{t\rightarrow 0}||\mathcal{T}_t- I||=0,$ then $\mathcal{T}$ has a generator $\mathcal{L}=\lim_{t\rightarrow\infty}(\mathcal{T}_t-I)/t$ (see \cite{Lind}), which is a bounded operator on $\mathcal{I}_1(\mathcal{H}),$ also known as \textbf{Lindblad operator}.\index{Lindblad operator}

The set of density operators on a Hilbert space $\mathcal{K}$ will be denoted by
\begin{equation}\label{states2}\nonumber
\mathcal{S}({\mathcal{K}}):=\{\rho\in\mathcal{I}_1(\mathcal{K}),\;\rho\geq 0,\;\mbox{Tr}(\rho)=1\}
\end{equation}
and the set of block-diagonal density operators of $\mathcal{H}$ by
\begin{equation}\label{blockstates}\nonumber
\mathcal{D}:=\left\{\rho\in\mathcal{S}(\mathcal{H}):\;\rho=\sum_{i\in V}\rho(i)\otimes\ket{i}\bra{i}\right\}.
\end{equation}
This means that if $\rho\in \mathcal{D},$ then $\rho(i)\in \mathcal{I}_1(\mathfrak{h}_i),\;\rho(i)\geq 0$ (positive semidefinite) and $\sum_{i\in V}\mathrm{Tr}(\rho(i))=1.$

Let $[A,B]\equiv AB-BA$ denote the commutator between $A$ and $B$ and $\{A,B\}\equiv AB+BA$ the anti-commutator between $A$ and $B.$

\begin{defi}
Let $V$ be a finite or countable infinite set and $\mathcal{H}$ be a Hilbert space of the form \eqref{H}. A \textbf{Continuous-time Open Quantum Walk}\index{Continuous-time Open Quantum Random Walk CTOQW} (CTOQW) in $V$ is an uniformly continuous QMS on $\mathcal{I}_1(\mathcal{H})$ with Lindblad
operator of the form
\begin{eqnarray}
\nonumber  \mathcal{\mathcal{L}}:\mathcal{I}_1(\mathcal{H}) &\rightarrow &  \mathcal{I}_1(\mathcal{H})\\\label{Lind}
  \rho &\mapsto & -i[\textbf{H},\rho]+\sum_{i,j\in V}\left(S_i^j\rho S_i^{j^*}-\frac{1}{2}\{S_i^{j*}S_i^j,\rho\}\right).
\end{eqnarray}
We can write $S_i^j=R_i^j\otimes\ket{j}\bra{i}$ for bounded operators $R_i^j\in \mathcal{B}(\mathfrak{h}_i,\mathfrak{h}_j),$ the bounded operators $\textbf{H}$ and $S_i^j$ on $\mathcal{H}$ are of the form $\textbf{H}=\sum_{i\in V}H_i\otimes\ket{i}\bra{i},$ $H_i$ is self-adjoint on $\mathfrak{h}_i,$ and $\sum_{i,j\in V}S_i^{j*}S_i^j$ converges in the strong sense.
\end{defi}

The generator $\mathcal{L}$ preserves $\mathcal{D}.$ That is, for an initial density $\rho=\sum_{i\in V}\rho(i)\otimes\ket{i}\bra{i}\in
\mathcal{D},$ $e^{t\mathcal{L}}(\rho)=\mathcal{T}_t(\rho)=\sum_{i\in V}\rho_t(i)\otimes\ket{i}\bra{i},\forall t\geq 0,$ with
$$
\frac{d}{dt}\rho_t(i)=-i[H_i,\rho_t(i)]+\sum_{j\in V}\left(R_j^i\rho_t(j) R_j^{i^*}-\frac{1}{2}\{R_i^{j*}R_i^j,\rho_t(i)\}\right).
$$

An alternative way to write \eqref{Lind} is given by equation (18.7) in \cite{bardet}:
\begin{equation*}\label{alternativeLindblad}
\mathcal{L}(\rho)=\sum_{i\in V}\left(G_i\rho(i)+\rho(i)G_i^*+\sum_{j\in V}R_j^i\rho(j) R_j^{i*}\right)\otimes\ket{i}\bra{i},
\end{equation*}
where
\begin{equation*}\label{Gi}
G_i=-iH_i-\frac{1}{2}\sum_{j\in V}R_i^{j*}R_i^j.
\end{equation*}
Since the operator $H_i$ is hermitian, we have $G_i+G_i^*=-\sum_{j\in V}R_i^{j*}R_i^j.$ The reader should compare this identity with the Zero Line-Sum Matrices (or $Q$-matrices), and make a connection with the classical Markov chains. For more details on this, see \cite[Proposition 3.9]{pele}.

Let us now explain the dynamics of a CTOQW: given a finite or countably infinite set of vertices $V,$ a CTOQW is a stochastic process evolving on a Hilbert space of the form \eqref{H}. The label $i\in V$ represents the position of the walker and when the walker is located at $i\in V,$ its internal state is encoded in $\mathfrak{h}_i,$ that is, $\mathfrak{h}_i$  describes the internal degrees of freedom of the walker when it is at site $i\in V.$

Starting the walk on site $\ket{i}$ with initial density operator $\rho\in\mathcal{S}(\mathfrak{h}_i)=\sum_{i\in V}\rho(i)\otimes\ket{i}\bra{i},$ the quantum measurement of the ``position" gives rise to a probability distribution $p_0$ on $V,$ such that
$$
p_0(i)=\mathbb{P}(\mbox{the quantum particle is in site}\;\ket{i})=\mathrm{Tr}(\rho(i)),
$$
and for evolution on time $t\geq 0,$
\begin{equation}\label{rho_t(i)}
p_t(i)=\mathbb{P}(\mbox{the quantum particle, at time }t,\mbox{ is in site}\;\ket{i})=\mathrm{Tr}(\rho_{t}(i)),
\end{equation}
where
$$
e^{t\mathcal{L}}(\rho)=\sum_{i\in V}\rho_t(i)\otimes\ket{i}\bra{i}.
$$

We say that the CTOQW is \textbf{semifinite}\index{semifinite!CTOQW} when $\dim(\mathfrak{h}_i)<\infty\;\forall i\in V.$ If in addition
$|V|<\infty,$ then the CTOQW is said to be \textbf{finite}.\index{finite CTOQW}

We follow \cite{bardet,pele} to discuss the quantum trajectory describing the indirect measurement of the position of a CTOQW of the more general form \eqref{Lind} in order to obtain probabilistic properties of this quantum system. For more general results of quantum trajectories, see \cite{KM}. So, let $(\Omega,\mathcal{F},(\mathcal{F}_t)_{t\geq 0},\mathbb{P})$ be a probability space where independent Poisson point processes $N^{ij},i,j\in,V,\; i\neq j$ ($N^{ii}=0$ by convention) on $\mathbb{R}^2$ are defined. The jump from site $i$ to site $j$ on the graph $V$ will be governed by these Poisson point processes.

\begin{defi}\label{XtDefi}
Consider a CTOQW with generator of the form \eqref{Lind} and an initial density operator $\mu=\sum_{i\in V}\rho(i)\otimes\ket{i}\bra{i}\in\mathcal{D}.$ The quantum trajectory describing the indirect measurement of the position of the CTOQW is the Markov chain represented by the density operators $(\mu_t)_{t\geq 0}$ such that $\mu_0=\rho_0\otimes\ket{X_0}\bra{X_0},$ where $X_0$ and $\rho_0$ are random variables with distribution
$$
\mathbb{P}\left((X_0,\rho)=\left(i,\frac{\rho(i)}{\mathrm{Tr}(\rho(i))}\right)\right)=\mathrm{Tr}\left(\rho(i)\right)\mbox{ for all }i\in V,
$$
and $\mu_t=:\rho_t\otimes\ket{X_t}\bra{X_t}$ satisfies the stochastic differential equation
\begin{equation}\label{muTrajectory}
\begin{split}
   \mu_t =& \mu_0+\int_{0}^{t}M(\mu_{s^-})ds \\
     +& \sum_{ij}\int_{0}^{t}\int_{\mathbb{R}}\left(\frac{S_{i}^{j}\mu_{s^-}S_{i}^{j*}}{\mathrm{Tr}(SS_{i}^{j*}\mu_{s^-}S_{i}^{j*})}-\mu_{s^-}\right)1_{0<y<\mathrm{Tr}(S_{i}^{j}\mu_{s^-}S_{i}^{j*})}N^{ij}(dy,ds)
\end{split}
\end{equation}
for all $t\geq 0,$ where
$$
M(u)=\mathcal{L}(u)-\sum_{ij}\left(S_{i}^{j}\mu S_{i}^{j*}-\mu\mathrm{Tr}(S_{i}^{j}\mu S_{i}^{j*})\right).
$$

Hence, for a fixed $\mu=\sum_{i}\rho(i)\otimes\ket{i}\bra{i}\in\mathcal{D},$
$$
M(\mu)=\sum_{i}\left(G_i\rho(i)+\rho(i)G_i^*-\rho(i)\mathrm{Tr}\left(G_i\rho(i)+\rho(i)G_i^*\right)\right)\otimes\ket{i}\bra{i}.
$$
\end{defi}

The evolution of the solution $\mu_t$ of \eqref{muTrajectory} is described as follows: suppose $X_0=i_0$ for some $i_0\in V$ and $\rho_0\in V(\mathfrak{h}_{i_0}).$ For all $t\geq 0,$ consider the solution
$$
\eta_t=\rho_0+\int_{0}^{t}\left(G_{i_0}\eta_s+\eta_sG_{i_0}^*-\eta_s\mathrm{Tr}\left(G_{i_0}\eta_s+\eta_sG_{i_0}^*\right)\right)ds,
$$
which is a density operator on $\mathfrak{h}_{i_0}.$ For $j\neq i_0,$ define
$$
T_1^j=\mathrm{inf}\{t\geq 0;N^{i_0,j}\left({u,y|0\leq u\leq t,0\leq y\leq \mathrm{Tr}(R_{i_0}^{j}\eta_u R_{i_0}^{j*})}\right)\geq 1\}.
$$
Since the random variables $T_1^j$ are mutually independent and nonatomic, we can define $T_1=\inf_{j\neq i_0}\{T_1^j\}$ once there exists a unique $j\in V$ such that $T_1^j=T_1.$ The random variable $T_1$ is said to be the \textbf{first jump time} of the CTOQW conditional on $X_0=i_0.$

The first jump time to site $\ket{j}$ is then denoted by $T_1^j$ and has distribution
$$
\mathbb{P}(T_1^j>\varepsilon)=e^{-\int_0^\varepsilon\mathrm{Tr}(R_{i_0}^{j}\eta_uR_{i_0}^{j*})du},
$$
thus
$$
\mathbb{P}(T_1\leq\varepsilon)\leq\varepsilon\sum_{j\neq i_0}\|R_{i_0}^{j*}R_{i_0}^{j}\|.
$$
The strongly convergence of $\sum_{ij}S_{i}^{j*}S_{i}^{j}$ implies that $\mathbb{P}(T_1>0)=1.$ Thereby, on $[0,T_1],$ we can define the solution $(X_t,\rho_t)_{t\geq 0}$ as
\begin{eqnarray}
\nonumber (X_t,\rho_t)  &=& (i_0,\eta_t)\mbox{ for }t\in[0,T_1)\mbox{ and } \\
\nonumber (X_{T_1},\rho_{T_1})  &=& \left(j,\frac{R_{i}^{j}\eta_{T_1-}R_{i}^{j*}}{\mathrm{Tr}(R_{i}^{j}\eta_{T_1-}R_{i}^{j*})}\right)\;\mbox{ if }T_1=T_1^j.
\end{eqnarray}

Now we solve
\begin{equation*}\label{eta_t defi}
\eta_t=\rho_{T_1}+\int_{0}^{t}\left(G_{j}\eta_s+\eta_sG_{j}^*-\eta_s\mathrm{Tr}\left(G_{j}\eta_s+\eta_sG_{j}^*\right)\right)ds
\end{equation*}
and then obtain the second jump time $T_2.$ So on we obtain an increasing sequence of jumps $(T_n)_n$ with $\lim_{n\rightarrow\infty}T_n=\infty$ almost surely (see section 18.2.3 of \cite{bardet} for more details). This means that the walk does not explode, thus the walker has a finite number of jumps in a finite interval. For details concerning explosions of classical Markov chains, see \cite[Section 2.2]{norris}.

The CTOQWs of our interest in this work will be defined now:

\begin{defi}
Consider a CTOQW on $V=\mathbb{Z}$.
\begin{enumerate}
  \item It is called \textbf{homogeneous} if
$$R_{j}^{i}=R_{j+k}^{i+k}$$
for every $i,j,k\in\mathbb{Z}.$

\medskip

  \item If it is homogeneous and there exist operators $C,A,H$ such that
$$R_{i+1}^{i}=C,\quad R_{i}^{i+1}=A,\quad H_i=H\;\; \forall i\in V\quad and \quad R_{j}^{i}=0 \mbox{ for } |j-i|\neq 1,$$
   then we say that the CTOQW is \textbf{induced by a coin} $(C,A)_H.$
\end{enumerate}
\end{defi}

When a CTOQW is induced by a coin $(C,A)_H,$ then we can let $\mathfrak{h}_i=\mathfrak{h}$ for all $i\in \mathbb{Z}.$ If $\dim(\mathfrak{h})=d,$ then $C,A$ and $H$ can be represented by square matrices of order $d$ and we say that $(C,A)_H$ is a coin of dimension $d.$

A CTOQW with generator $\mathcal{L}$ is \textbf{irreducible} when, for all $X\in\mathcal{I}_1(\mathcal{H})$ with $X\geq 0$ and $X\neq 0,$ there exists $t>0$ such that $e^{t\mathcal{L}}(X)>0.$

\subsection{Recurrence and Transience for CTOQWs} This section is devoted to introducing concepts of recurrence of CTOQWs following some classical definitions, but with the introduction of the initial quantum state, which is a density operator.

Analogous to the classical walk (see \cite{ander}), we pick $\delta>0$ in order to discretize a CTOQW by a process $\{X(n\delta),n\geq 0\}$ having one-step transition probability $p_{ji;\rho}(\delta)$ (thus it has $n$ step transition probabilities $p_{ji;\rho}(n\delta)$). This process will be called the $\delta$-\textbf{skeleton} of $\{X(t),t\geq 0\}$ and one can check that it performs a discrete Open Quantum Walk.

With this notation, $p_{ji;\rho}(t)$ denotes the probability of being at site $j$ at time $t,$ given that we started at site $i,$ with initial density $\rho$ concentrated at $i,$ that is,
\begin{equation*}\label{ctoqwprobmat}
p_{ji;\rho}(t)=p_{t}(\rho\otimes |i\rangle \to |j\rangle)=\mathrm{\mathrm{Tr}}(\rho_t(j)\otimes\ket{j}\bra{j})=\mathrm{\mathrm{Tr}}\left(e^{t\mathcal{L}}(\rho\otimes\ket{i}\bra{i})(I\otimes\ket{j}\bra{j})\right).
\end{equation*}
Therefore, the dynamics starts with a density operator $\rho$ concentrated at some vertex $\ket{i},$ takes the evolution up to time ($t$) through the exponential of the Lindblad operator $\mathcal{L},$ producing a new density operator
$$\rho_t=\sum_k\rho_t(k)\otimes\ket{k}\bra{k}=e^{t\mathcal{L}}(\rho\otimes\ket{i}\bra{i}),\quad \mathrm{Tr}\left(\sum_k\rho_t(k)\right)=1,$$
and then we project $\rho_t$ onto the subspace generated by vertex $\ket{j}.$ Note that
$$
e^{t\mathcal{L}}(\rho\otimes\ket{i}\bra{i})(I\otimes\ket{j}\bra{j})=\rho_t(j)\otimes\ket{j}\bra{j},
$$
which represents the data concentrated at vertex $\ket{j}$ at time $t.$

Consider a semifinite CTOQW on some set of vertices $V$, let $i\in V,\rho\in \mathcal{S}(\mathfrak{h}_i)$ and $\delta>0.$  We say that a vertex $i$ is
\begin{itemize}
  \item $\rho$-\textbf{recurrent}\index{$\rho$-recurret vertex} if
  \begin{equation*}\label{prec}
    \int_{0}^{\infty}p_{ii;\rho}(t)dt=\infty.
  \end{equation*}
Otherwise, $i$ is said to be $\rho$-\textbf{transient};\index{$\rho$-transient vertex}
  \item $\rho$-\textbf{SJK-recurrent}\footnote{The notion of SJK-recurrence in the discrete-time unitary setting is described on
\cite{stef}. SJK-recurrence is named after the initials of the authors of that work.}\index{$\rho$-SJK-recurrent vertex} in the
$\delta$-skeleton if
\begin{equation*}\label{psjkrec}
    \sum_{n=0}^{\infty}p_{ii;\rho}(n\delta)=\infty.
  \end{equation*}
Otherwise, $i$ is said to be $\rho$-\textbf{SJK-transient} in the $\delta$-skeleton;\index{$\rho$-SJK-transient vertex}
  \item \textbf{recurrent},\index{recurrent vertex} if $i$ is $\rho-$recurrent for all $\rho\in S_{\mathfrak{h}_i};$
  \item \textbf{transient},\index{transient vertex} if $i$ is $\rho-$transient for all $\rho\in S_{\mathfrak{h}_i};$
  \item \textbf{SJK-recurrent},\index{SJK-recurrent vertex} if $i$ is $\rho$-SJK-recurrent in the $\delta$-skeleton for all $\rho\in
      S_{\mathfrak{h}_i};$
  \item \textbf{SJK-transient},\index{SJK-transient vertex} if $i$ is $\rho$-SJK-transient in the $\delta$-skeleton for all $\rho\in
      S_{\mathfrak{h}_i}.$
\end{itemize}

\begin{REM}
Further in this section, assuming that the walk is semifinite, we shall show that a vertex is $\rho$-recurrent in the $\delta$-skeleton if, and only if, it is $\rho$-recurrent in the $\delta'$-skeleton for any $\delta'>0.$ So, the definitions of $SJK$-recurrence and
$SJK$-transience are consistent.
\end{REM}

\begin{defi}
A CTOQW is said to be:
\begin{itemize}
  \item \textbf{recurrent}\index{recurrent CTOQW} if every vertex is recurrent;
  \item \textbf{transient}\index{transient CTOQW} if every vertex is transient;
  \item \textbf{SJK-recurrent}\index{SJK-recurrent CTOQW} if every vertex is SJK-recurrent;
  \item \textbf{SJK-transient}\index{SJK-transient CTOQW} if every vertex is SJK-transient.
\end{itemize}
\end{defi}

When the CTOQW is semifinite and irreducible, we have the following dichotomy due to \cite{bardet}:
\begin{prop}\label{BardetDichotomy}
Consider a semifinite irreducible CTOQW. We are in one (and only one) of the following situations:
\begin{itemize}
  \item Every vertex is recurrent;
  \item Every vertex is transient.
\end{itemize}
\end{prop}

If a CTOQW is induced by a coin of finite dimension, then we can determine the set of densities operators in which the CTOQW is transient by analyzing the common invariant subspaces of the components of the coin. Let us explain this in the following remark.

\begin{remark}\label{RemInvSub}
Let us consider a CTOQW induced by a coin $(C,A)_H$ of dimension $d<\infty,$ and assume that $(C,A)_H$ is $\rho$-transient. We suppose that the walk starts on vertex $\ket{0}$ with initial density operator $\rho,$ then after a jump on instant $T_1,$ the new density is
$$\rho_{T_1}=\frac{A\rho A^*}{\mathrm{Tr}(A\rho A^*)}\;\mbox{ or }\;\rho_{T_1}=\frac{C\rho C^*}{\mathrm{Tr}(C\rho C^*)}.$$

If we consider a new CTOQW starting on vertex $\ket{-1}$ or $\ket{1}$ with $\rho_{T_1}$ as given above, then the walk must be $\rho_{T_1}$-transient, by the homogeneity in space of the walk. So on we see that $(C,A)_H$ is $\rho_{T_n}$ transient for every $n=1,2,\ldots,$ where $\rho_{T_n}$ is the $n$-th jump. Therefore the set of densities $\{\rho_{T_n}\}_n$ must be contained in a subset $\mathcal{S}(\mathcal{Y})\subseteq \mathcal{D},$ where $\mathcal{Y}$ is a common invariant subspace of both $C$ and $A$.

In particular, if $(C,A)_H$ is a coin of dimension $2,$ then $\mathcal{Y}$ is or a common eigenspace of $A$ and $C$ or one of the trivial common subspaces $\{0\},\;\mathbb{C}^2.$
\end{remark}

\subsection{Properties of Probability Transitions }

In this subsection, we present generalizations of some results about Markov Chains described previously. For instance, the Chapman-Kolmogorov Identity, which arises from the Markov Property, however, the quantum model is non-Markovian, hence we obtain an equation that seems like Chapman-Kolmogorov Identity, but with a perturbation on the initial quantum state.

The proofs of the results presented in this subsection will be given in the Appendices (Section 6).

\begin{prop}\label{lard} Consider a CTOQW in $V,$ $\alpha,\beta\geq 0,$ and let $j,i\in V$, then
\begin{equation*}\label{chapquantum}
p_{ji;\rho}(\alpha+\beta)=\sum_{k}p_{jk;\rho'_{ki}(\beta)}(\alpha)p_{ki;\rho}(\beta),
\end{equation*}
where
$$
\rho'_{ki}(\beta)=\frac{e^{\beta\mathcal{L}}(\rho\otimes\ket{i}\bra{i})P_k}{\mbox{Tr}(e^{\beta\mathcal{L}}(\rho\otimes\ket{i}\bra{i})P_k)},\quad k\in V.
$$
\end{prop}

Letting $\dim(\mathfrak{h}_i)=1\;\forall i\in V$ gives $\rho'_{ki}(\beta)=1\in \mathbb{C},\;\forall k,i\in V.$ Therefore, this formula gives the Chapman-Kolmogorov Identity for continuous-time Markov chains.

To show equivalence between recurrence and SJK-recurrence, we will show that the function $g(\rho,s)=p_{ji;\rho}(s)$ is jointly continuous under the variables $s\in [0,\infty)$ and $\rho\in\mathcal{S}(\mathfrak{h}_i),$ which is a result that is not necessary to show this equivalence for classical Markov chains.

\begin{prop}\label{jointlycontinuous}Consider a CTOQW in $V$, let $i,j\in V$ and denote $W_i:=\mathcal{S}(\mathfrak{h}_i)\times[0,+\infty).$ The function $g:W_i\rightarrow
[0,1]$ defined by $g(\rho,s)=p_{ji;\rho}(s)$ is jointly continuous on $W_i.$
\end{prop}

The following proposition gives properties of the transition functions of CTOQWs. They have fundamental importance to the next definitions and results since they give sufficient conditions to the transition functions to be strictly positive for $t$ sufficiently large. Items (1), (2), (3) generalize Proposition 1.3 of \cite{ander} and (4) associate the positivity of $p_{ji;\rho}(t)$ with all the initial densities operators.

\begin{prop}\label{cont and min}
Consider a CTOQW in $V$ and let $i\in V.$ The following assertions hold.
\begin{enumerate}
  \item For all $\rho\in\mathcal{S}(\mathfrak{h}_i)\mbox{ and }t\geq 0,$ $p_{ii;\rho}(t)>0;$
  \item If $p_{ij;\rho}(t)>0$ for some $t>0,$ then $p_{ij;\rho}(s)>0\;\forall s\geq t;$
  \item If $p_{ii;\rho}(t)=1$ for some $t>0,$ then $p_{ii;\rho}(s)=1,\forall s\in[0,t];$
  \item If $\delta>0,$ $\dim(\mathfrak{h}_i)<\infty$ and there exists $t_0\in[0,\delta]$ such that $p_{ji;\rho}(t_0)>0,$ then the minimum
      $N_{ji}:=\min\{p_{ji;\rho}(s):\rho\in \mathcal{S}(\mathfrak{h}_i)\mbox{ and }s\in[t_0,\delta]\}$ is attained on $(0,1]$.
\end{enumerate}
\end{prop}

We denote for $i,k\in V$ and $\delta>0$ the following values
\begin{equation*}\label{Xi}
\Xi_{ki}^\delta(\rho)=\sum_{n=0}^{\infty}p_{ki;\rho}(n\delta)
\end{equation*}
and
\begin{equation*}\label{iX}
\mathbb{E}_{i,\rho}(n_k)=\int_{0}^{\infty}p_{ki;\rho}(t)dt.
\end{equation*}

Those values are associated with the following Theorem. Proof of the classical result can be found in \cite[Proposition 5.1.1]{ander}.

\begin{teo}\label{equiv} Consider a CTOQW in $V,$ $\delta>0$ and $i,j\in V.$ If $\dim(\mathfrak{h}_j)<\infty,$ then
\begin{equation*}\label{equiv1}
\Xi_{ji}^\delta(\rho)=+\infty\quad\Leftrightarrow\quad\mathbb{E}_{i,\rho}(n_j)=+\infty.
\end{equation*}
In particular, $j$ is $\rho$-recurrent if, and only if, $j$ is $\rho$-SJK-recurrent in the $\delta$-skeleton.
\end{teo}

Now we have an equivalence between the CTOQW and its discretized random walk with $n$-step transition probabilities in the $\delta$-skeleton. To finish this section, we give some results about site recurrence in terms of the initial density operators.

\begin{prop}\label{discussrec}
Consider a CTOQW in $V$, $i\in V,$ $\dim(\mathfrak{h}_i)=n<\infty,\;\tilde{\rho}\in\mathcal{S}(\mathfrak{h}_i)$ and suppose that $i$ is
$\tilde{\rho}$-recurrent.
\begin{enumerate}
  \item For any faithful $\rho\in\mathcal{S}(\mathfrak{h}_i),$ $i$ is $\rho$-recurrent;
  \item If $\rho\in\mathcal{S}(\mathfrak{h}_i)$ and there exists $\delta\geq 0$ such that $\rho'_{ii}(\delta)$ is faithful, then $i$ is
      $\rho$-recurrent;
  \item If $n\geq 2,$ there is a non-faithful density $\rho$ on $\mathcal{S}(\mathfrak{h}_i)$ in which $i$ is $\rho$-recurrent;
  \item If $n\geq 2$, then the non-faithful density $\rho$ on item (3) can be assumed to be pure;
  \item If $n=2,$ then $i$ is $\rho$-transient for at most one density $\rho$.
\end{enumerate}
\end{prop}

\begin{REM}
By contraposition, we get by the first item of the Proposition \ref{discussrec} that if $i\in V,$ $\dim(\mathfrak{h}_i)=n<\infty$ and
$\rho\in\mathcal{S}(\mathfrak{h}_i)$ is faithful with $i$ being $\rho$-transient, then $i$ is $\rho'$-transient for any $\rho'$.
\end{REM}

\section{Recurrence criteria for CTOQWs induced by a finite coin}

Now we consider the special generator of CTOQWs of the form
\begin{equation}\label{newLgen}
\mathcal{L}(\mu)=-i[\textbf{H}\otimes I,\mu]+\sum_{i\in \mathbb{Z}^d}\sum_{r=1}^{2d}\left(B_i^r\mu B_i^{r*}-\frac{1}{2}\left\{B_i^{r*}B_i^r,\mu\right\}\right),
\end{equation}
where $\{e_1,\ldots,e_r\}$ is the canonical basis of $\mathbb{Z}^d$ and we set $e_0=0_d,\;e_{d+r}=-e_{r}$ for all $r\in\{1,\ldots,d\},$ and $B_i^r=D_r\otimes\ket{i+e_r}\bra{i}.$

The \textbf{auxiliary Lindblad operator} $\mathbb{L}:\mathcal{B}(\mathfrak{h})\rightarrow \mathcal{B}(\mathfrak{h})$ of a CTOQW with generator of the form \eqref{newLgen} is defined by
$$
\mathbb{L}(\rho)=G_0\rho+\rho G_0^*+\sum_{r=1}^{2d}D_r\rho D_r^*,
$$
where
$$G_0=-iH-\frac{1}{2}\sum_{r=1}^{2d}D_r^*D_r.$$
Given such an auxiliary Lindblad operator, we can consider the hypothesis
\begin{itemize}
  \item (H1): there exists a unique density operator $\rho_{inv}\in \mathcal{S}(\mathfrak{h})$ such that $\mathbb{L}(\rho_{inv})=0.$
\end{itemize}

Concerning CTOQWs induced by a coin, we can imagine the concept of recurrence with a scale of two pans, where there are many weights and they depend on the operators $A$ and $C$. In this situation, the CTOQW is recurrent if and only if the scale is in a state of equilibrium, thus one may conclude that there is a value depending on $A$ and $C$ that guarantees this equilibrium. As will be shown later, when $(H1)$ holds, this value is
\begin{equation}\label{md1}
m=\mathrm{Tr}(A\rho_{inv}A^*)-\mathrm{Tr}(C\rho_{inv}C^*),
\end{equation}
thus the equilibrium indeed depends on $A$ and $C$ and also on $\rho_{inv}$ (and therefore it also depends on the Hamiltonian operator). In a more general way, we let
$$m=\sum_{r=1}^{2d}\mathrm{Tr}\left(D_r\rho_{inv}D_r^*\right)e_r.$$

As discussed above, if $V=\mathbb{Z}$ ($d=1$), we let $D_1=A,\;D_2=C,$ $m$ becomes the value of Equation \eqref{md1} {and the auxiliary Lindblad operator has the form
\begin{equation}\label{LforCoin}
\mathbb{L}(\rho)=-iH\rho+i\rho H-\frac{1}{2}\left(C^*C+A^*A\right)\rho-\frac{1}{2}\rho\left(C^*C+A^*A\right)+C\rho C^*+A\rho A^*.
\end{equation}

\begin{lema}[\cite{bri}]
For all $u\in\mathbb{R}^d,$ the equation
\begin{equation}\label{Lestrela}
\mathcal{L}^*(J_u)=-\left(\sum_{r=1}^{2d}(e_r.u)D_r^*D_r-(m.u)I\right)
\end{equation}
admits a solution and the difference between any couple of solutions of \eqref{Lestrela} is a multiple of the identity.
\end{lema}

The following Theorem is a continuous-time version of the Law of Large Numbers given in \cite[Theorem 5.2]{AttalCTL}, and it is a consequence of the results of \cite{bri}.

\begin{teo}[Law of Large Numbers for CTOQWs]\label{LawofLargeNumbers}
Let $\left(\rho_t,X_t\right)_{t\geq 0}$ be the Markov process on Definition \ref{XtDefi} and assume that the generator is of the form \eqref{newLgen}. If (H1) holds, then
\begin{equation*}\label{LLN}
\frac{X_t}{t}\rightarrow m\quad \mbox{a.s}.
\end{equation*}
\begin{proof}
We have the following assertions from \cite[Theorem 3.0.9]{bri}:
\begin{enumerate}
  \item The process $(M_t)_{t\geq 0}$ defined by
$$M_t=\mathrm{Tr}(\rho_tJ_u)-\mathrm{Tr}(\rho_0J_u)+X_t.u-X_0.u-(m.u)t $$
is a martingale with respect to the filtration associated with $(\rho_t,X_t)_{t\geq 0};$
  \item $\Delta M_s$ is bounded independently of $s;$
  \item $\mathrm{Tr}(\rho_tJ_u)-\mathrm{Tr}(\rho_0J_u)-X_0.u$ is bounded independently of $t.$
\end{enumerate}

Using items (1) and (2), Azuma's inequality and the Borel Cantelli lemma, we get $M_s/s\rightarrow 0$ a.s. Now we use item (3) to obtain
$$\frac{X_t}{t}\rightarrow m\quad \mbox{a.s.}$$
\end{proof}
\end{teo}

The following Lemma is adapted from the discrete case (see \cite[Lemma 6]{JL2}), however, this continuous-time version gives a stronger result in terms of invariant subspaces.

\begin{lema}\label{lemmaMax}
Consider a homogeneous CTOQW with $V=\mathbb{Z},$ let $(X_t,\rho_t)_{t\geq 0}$ be the Markov chain given on Definition \ref{XtDefi} and assume that $\mathcal{Y}$ is a subspace of $\mathbb{C}^d$ which is invariant for each $R_i^j.$ Let $\varepsilon>0$ and $\omega>1$ an integer, then for every $\rho\in\mathcal{D}:=\mathcal{D}(\mathcal{Y}),$ we have
$$
\int_{0}^{\infty}\mathbb{P}_{0,\rho}\left(|X_t|<\omega\varepsilon\right)dt\leq 2\omega.\max_{\sigma\in\mathcal{D}}\int_{0}^{\infty}\mathbb{P}_{0,\sigma}\left(|X_t|<\varepsilon\right) dt.
$$
\begin{proof}
We write $(-\omega,\omega)\subset \cup_{k=-\omega}^{\omega-1}[k,k+1)$ to get the inequality
$$
\int_{0}^{\infty}\mathbb{P}_{0,\rho}\left(|X_t|<\omega\varepsilon\right)dt\leq\int_{0}^{\infty}\sum_{k=-\omega}^{\omega-1}\mathbb{P}_{0,\rho}\left(k\varepsilon\leq X_t<(k+1)\varepsilon\right)dt.
$$

Denote $T_k=\inf\{y\geq 0:k\varepsilon\leq X_y<(k+1)\varepsilon\}$ and $$\zeta_{\varepsilon,k}^\rho=\int_{0}^{\infty}\mathbb{P}_{0,\rho}\left(k\varepsilon\leq X_t<(k+1)\varepsilon\right)dt,$$
then we obtain the identity
$$\zeta_{\varepsilon,k}^\rho=\int_{0}^{\infty}\int_{0}^{\infty}\mathbb{P}_{0,\rho}\left(k\varepsilon\leq X_t<(k+1)\varepsilon,T_k=y\right)dydt.$$
For $T_k=y,$ we have $k\varepsilon\leq X_y<(k+1)\varepsilon,$ thus $-(k+1)\varepsilon<-X_y\leq -k\varepsilon$ and $-\varepsilon<X_t-X_y<\varepsilon,$ thereby Fubini's Theorem gives
\begin{equation}\nonumber
\begin{split}
\zeta_{\varepsilon,k}^\rho=&\int_{0}^{\infty}\int_{0}^{\infty}\mathbb{P}_{0,\rho}\left(|X_t-X_y|<\varepsilon,T_k=y\right)dydt  \\
     =& \int_{0}^{\infty}\int_{y}^{\infty}\mathbb{P}_{0,\rho}\left(|X_t-X_y|<\varepsilon,T_k=y\right)dtdy\\
     =&\int_{0}^{\infty}\mathbb{P}_{0,\rho}\left(T_k=y\right)\int_{y}^{\infty}\mathbb{P}_{0,\rho}\left(|X_t-X_y|<\varepsilon\right)dtdy,
\end{split}
\end{equation}
where the last equality follows from independence.

Denote $\mathbb{P}_{0,\rho}^{k,y}=\mathbb{P}_{0,\rho}\left(T_k=y\right).$  Since $\mathcal{Y}$ is invariant by all $R_i^j,$ the quantum trajectory never leaves $\mathcal{D},$ thus a summation over $\mathcal{D}$ and $\mathbb{Z}$ and an application of Fubini's Theorem give
\begin{equation}\nonumber
\begin{split}
\zeta_{\varepsilon,k}^\rho=& \int_{0}^{\infty}\mathbb{P}_{0,\rho}^{k,y}\int_{y}^{\infty}\mathbb{P}_{0,\rho}\left(|X_t-X_y|<\varepsilon\right)dtdy \\
     =&\int_{0}^{\infty}\mathbb{P}_{0,\rho}^{k,y}\int_{y}^{\infty}\sum_{\sigma\in\mathcal{D},j\in\mathbb{Z}}\mathbb{P}_{0,\rho}\left(|X_t-X_y|<\varepsilon,(X_y,\rho_y)=(j,\sigma)\right)dtdy \\
     =&\int_{0}^{\infty}\mathbb{P}_{0,\rho}^{k,y}\sum_{\sigma\in\mathcal{D},j\in\mathbb{Z}}\int_{y}^{\infty}\mathbb{P}_{0,\rho}\left(|X_t-X_y|<\varepsilon|(X_y,\rho_y)=(j,\sigma)\right)\mathbb{P}_{0,\rho}\left((X_y,\rho_y)=(j,\sigma)\right)dtdy.
\end{split}
\end{equation}
The summation over $\mathcal{D}$ is valid because $\mathbb{P}_{0,\rho}\left((X_y,\rho_y)=(j,\sigma)\right)$ is supported on a finite set as a function of $\sigma.$

By homogeneity on space and the change of variables $s=t-y$, we get
\begin{equation}\nonumber
\begin{split}
\zeta_{\varepsilon,k}^\rho=&\int_{0}^{\infty}\mathbb{P}_{0,\rho}^{k,y}\sum_{\sigma\in\mathcal{D},j\in\mathbb{Z}}
\int_{y}^{\infty}\mathbb{P}_{j,\sigma}\left(|X_{t-y}-X_0|<\varepsilon\right)dt\mathbb{P}_{0,\rho}\left((X_y,\rho_y)=(j,\sigma)\right)dy \\
=&\int_{0}^{\infty}\mathbb{P}_{0,\rho}^{k,y}\sum_{\sigma\in\mathcal{D},j\in\mathbb{Z}}\int_{0}^{\infty}\mathbb{P}_{j,\sigma}
\left(|X_{s}-j|<\varepsilon\right)ds\mathbb{P}_{0,\rho}\left((X_y,\rho_y)=(j,\sigma)\right)dy \\
=&\int_{0}^{\infty}\mathbb{P}_{0,\rho}^{k,y}\sum_{\sigma\in\mathcal{D}}\int_{0}^{\infty}\mathbb{P}_{0,\sigma}
\left(|X_{s}|<\varepsilon\right)ds\sum_{j\in\mathbb{Z}}\mathbb{P}_{0,\rho}\left((X_y,\rho_y)=(j,\sigma)\right)dy \\
\leq&\int_{0}^{\infty}\mathbb{P}_{0,\rho}^{k,y}\max_{\tau\in\mathcal{D}}\int_{0}^{\infty}
\mathbb{P}_{0,\tau}\left(|X_{s}|<\varepsilon\right)ds\sum_{\sigma\in\mathcal{D},j\in\mathbb{Z}}\mathbb{P}_{0,\rho}\left((X_y,\rho_y)=(j,\sigma)\right)dy\\
=&\int_{0}^{\infty}\mathbb{P}_{0,\rho}^{k,y}dy\max_{\tau\in\mathcal{D}}\int_{0}^{\infty}
\mathbb{P}_{0,\tau}\left(|X_{s}|<\varepsilon\right)ds\\
=&\max_{\tau\in\mathcal{D}}\int_{0}^{\infty}
\mathbb{P}_{0,\tau}\left(|X_{s}|<\varepsilon\right)ds.
\end{split}
\end{equation}

Therefore,
\begin{equation}\nonumber
\begin{split}
\int_{0}^{\infty}\mathbb{P}_{0,\rho}\left(|X_{t}|<\omega\varepsilon\right)dt\leq&\sum_{k=-\omega}^{\omega-1}\zeta_{\varepsilon,k}^\rho\leq
 \sum_{k=-\omega}^{\omega-1}\max_{\tau\in\mathcal{D}}\int_{0}^{\infty}\mathbb{P}_{0,\tau}\left(|X_{s}|<\varepsilon\right)ds\\
=&2\omega.\max_{\tau\in\mathcal{D}}\int_{0}^{\infty}\mathbb{P}_{0,\tau}\left(|X_{s}|<\varepsilon\right)ds.
\end{split}
\end{equation}
\end{proof}
\end{lema}

Some properties regarding the recurrence of CTOQWs induced by coins will be presented below as applications of the Law of Large Numbers and the Lemma above.

\begin{teo}[Chung-Fuchs Theorem for CTOQWs]\label{ChungFuchs} Consider a CTOQW induced by a coin $(C,A)_H$ of dimension $d<\infty$ with
$$
\frac{X_t}{t}\rightarrow 0\quad\mbox{ in probability}.
$$

If $\mathcal{Y}$ is a subspace of $\mathbb{C}^d$ that is invariant for $C$ and $A,$ then there exists $\sigma\in \mathcal{D}=\mathcal{D}(\mathcal{Y})$ such that vertex $\ket{0}$ is $\sigma$-recurrent.
\begin{proof}
Firstly we denote $u_t^\rho(x)=\mathbb{P}_{0,\rho}\left(|X_{t}|<x\right)$ and by $\sigma$ the density that attains the maximum
$$\int_{0}^{\infty}u_t^\sigma(1)=\max_{\rho\in \mathcal{D}}\int_{0}^{\infty}u_t^\rho(1).$$
By lemma \ref{lemmaMax},
$$
2\omega.\int_{0}^{\infty}u_t^\sigma(1)dt\geq \int_{0}^{\infty}u_t^\rho(\omega)dt,\quad 1<\omega\in\mathbb{Z},\quad \rho\in\mathcal{D}.
$$

Let $A>0$ an integer, then
\begin{equation}\label{EqA(1)}
\begin{split}
\int_{0}^{\infty}p_{00;\sigma}(t)dt=&\; \int_{0}^{\infty}u_t^\sigma(1)\geq \frac{1}{2\omega}\int_{0}^{\infty}u_t^\rho(\omega) dt \\
\geq&\;  \frac{1}{2\omega}\int_{0}^{A\omega}u_t^\rho(\omega)dt\geq \frac{1}{2\omega}\int_{0}^{A\omega}u_t^\rho(t/A)dt,
\end{split}
\end{equation}
since $u_t^\rho(x)$ is a non-negative increasing function on $x$.

By the Weak Law hypothesis, we have
$$
\lim_{\omega\rightarrow \infty}\frac{1}{2\omega}\int_{0}^{\omega A}u_t^\rho(t/A)dt=\frac{A}{2}\lim_{\omega\rightarrow\infty}\dfrac{\displaystyle\int_{0}^{\omega A}u_t^\rho(t/A)}{\omega A}=\frac{A}{2},
$$
thus Equation \eqref{EqA(1)} gives
$$
\int_{0}^{\infty}p_{00;\sigma}(t)dt\geq \frac{A}{2}
$$
for every integer $A>0,$ from where we conclude that $\ket{0}$ is $\sigma$-recurrent.
\end{proof}
\end{teo}

Those results bring us to the main result of this paper:

\begin{teo}[Recurrence criteria for CTOQWs satisfying (H1)]\label{corR}
Consider a CTOQW induced by a coin $(C,A)_H.$ If condition (H1) holds, then
\begin{enumerate}
  \item $\mathrm{Tr}(A\rho_{inv}A^*)=\mathrm{Tr}(C\rho_{inv}C^*)\Rightarrow\;(C,A)_H$ is recurrent;
  \item $\mathrm{Tr}(A\rho_{inv}A^*)\neq\mathrm{Tr}(C\rho_{inv}C^*)\Rightarrow\;(C,A)_H$ is transient.
\end{enumerate}
\begin{proof}
Let $m=\mathrm{Tr}(A\rho_{inv}A^*)-\mathrm{Tr}(C\rho_{inv}C^*).$ If $m=0,$ we have by Theorem \ref{LawofLargeNumbers} that $X_t/t\rightarrow 0$ and thus Theorem \ref{ChungFuchs} assures that $\ket{0}$ is $\sigma$-recurrent for some $\sigma\in \mathcal{Y},$ where $\mathcal{Y}$ is any common invariant subspace of $C$ and $A.$ If $(C,A)_H$ was $\rho$-transient for some $\rho,$ then it would be transient with respect to all densities of $\mathcal{S}(\mathcal{X}),$ where $\mathcal{X}$ is a common invariant subspace for $C$ and $A,$ which is a contradiction.

On the other hand, if $m\neq 0,$ we have $X_t/t\rightarrow m\neq 0$ by Theorem \ref{LawofLargeNumbers}, thus $X_t$ must be unbounded on probability, that is,
$$
\mathbb{P}_{i,\rho}\left(X_t=0\;\mbox {i.o.}\right)=0,
$$
for any $\rho.$ Therefore the number of visits to $\ket{0}$ is finite for any initial density operator, thus $(C,A)_H$ is transient.
\end{proof}
\end{teo}

We remark that, fixed a coin $(C,A)_H,$ the CTOQW induced by this coin may be recurrent or transient according to the Hamiltonian operator since it changes the value of the stationary state $\rho_{inv}$ appearing in Theorem \ref{corR} (See Example \ref{Ex2} below).

\section{The $\mathbb{C}^2$ case}

This section is restricted to CTOQWs induced by coins of dimension 2. This lower dimension allows us to give a complete recurrence criterion for this kind of CTOQW in terms of its generator.

\begin{prop}\label{prop2inv}
Consider a CTOQW induced by a coin $(C,A)_H$ of dimension 2. If $\mathbb{L}$ has more than one invariant state, then $A$ and $C$ are both diagonal with respect to some orthonormal basis of $\mathbb{C}^2.$

Equivalently, $C$ and $A$ share two orthogonal eigenvectors.

\begin{proof}
We suppose that $\mathbb{L}$ has two distinct invariant states, then by \cite[Proposition 3]{SchWang}, there exist two proper orthogonal subspaces of $\mathbb{C}^2$ that are both invariant by $\mathbb{L}.$ Also as remarked in \cite{SchWang}, this implies that there exist two orthogonal subspaces of $\mathbb{C}^2$ that are both invariant for $C$ and $A.$ Those operators are acting in a Hilbert space of dimension 2, thereby they must share two orthogonal eigenvectors, and therefore they are normal operators, thus diagonal under the same basis.
\end{proof}
\end{prop}

\begin{lema}\label{lemmaH}
Consider a CTOQW induced by a coin $(C,A)_H$ of dimension 2, where the linear operators $C, A, H$ are of the form
\begin{equation*}
\begin{split}
C=& \begin{bmatrix}c_1 & 0 \\0 & c_2\end{bmatrix},\;
A=\begin{bmatrix}a_1 & 0 \\0 & a_2\end{bmatrix},\;
\quad a_j, \;c_j\in \mathbb{C},\quad j=1,2, \\
H=& \begin{bmatrix}h_1 & h_2 \\\overline{h_2} & h_3\end{bmatrix},\; 0\neq h_2\in \mathbb{C},\quad h_1,\;h_3\in\mathbb{R}.
\end{split}
\end{equation*}

If $\mathbb{L}$ has 2 distinct stationary states, then the Hamiltonian operator $H$ does not contribute with the recurrence of the walk and $a_1=a_2,\;c_1=c_2.$
\end{lema}
\begin{proof} Let us consider a CTOQW induced by a coin $(C,A)_H$ of dimension 2 with Lindblad generator $\mathcal{L},$ and
$$
\rho=\begin{bmatrix}\rho_{11} & \rho_{12} \\\overline{\rho_{12}} & 1-\rho_{11}\end{bmatrix},\quad\rho_{11}\in[0,1],\quad \rho_{12}\in \mathbb{C}, \quad
\mathbb{L}(\rho)=\begin{bmatrix}0 & 0 \\0 & 0\end{bmatrix},
$$
such that $|\rho_{12}|^2\leq \rho_{11}(1-\rho_{11}),$ that is, $\rho$ is a stationary state of $\mathbb{L}.$

In this case,
$$
\begin{bmatrix}0 & 0 \\0 & 0\end{bmatrix}=\mathbb{L}(\rho)=\begin{bmatrix}\tau_1 & \tau_2 \\\tau_3 & \tau_4\end{bmatrix},
$$
where
\begin{equation}\label{taus}
\begin{split}
\tau_1=&-2\mathrm{Im}(h_2\overline{\rho_{12}}),  \\
\tau_2=& \frac{\rho_{12}}{2}(-|a_1|^2-|a_2|^2-|c_1|^2-|c_2|^2+2a_1\overline{a_2}+2c_1\overline{c_2})+ih_2(2\rho_{11}-1)-i(h_1-h_3)\rho_{12}, \\
\tau_3=& \frac{\overline{\rho_{12}}}{2}(-|a_1|^2-|a_2|^2-|c_1|^2-|c_2|^2+2\overline{a_1}a_2+2\overline{c_1}c_2)-i\overline{h_2}(2\rho_{11}-1)+i(h_1-h_3)\overline{\rho_{12}}, \\
\tau_4=& 2i\mathrm{Im}(h_2\overline{\rho_{12}}).
\end{split}
\end{equation}

The values $\tau_j,\;j=1,2,3,4,$ were obtained with Equation \eqref{LforCoin}.

Note that $h_2\overline{\rho_{12}}\neq 0.$ Indeed, $h_2\neq 0$ by assumption and $\rho_{12}$ can not be null, otherwise we would have $\rho_{11}=1/2$ and $\rho=I/2$ would be the only stationary state of $\mathbb{L}.$

By the multiplication $h_2\tau_3=0,$ we get
\begin{equation*}
\begin{split}
0=    & \frac{h_2\overline{\rho_{12}}}{2}(-|a_1|^2-|a_2|^2-|c_1|^2-|c_2|^2+2\overline{a_1}a_2+2\overline{c_1}c_2)-i|h_2|^2(2\rho_{11}-1)+i(h_1-h_3)h_2\overline{\rho_{12}} \\
=& h_2\overline{\rho_{12}}\left(\frac{-|a_1|^2-|a_2|^2-|c_1|^2-|c_2|^2+2\overline{a_1}a_2+2\overline{c_1}c_2}{2}+i(h_1-h_3)\right)-i|h_2|^2(2\rho_{11}-1).
\end{split}
\end{equation*}
Recalling that $(h_1-h_3),\;|h_2|^2(2\rho_{11}-1)\in \mathbb{R},$ and that $h_2\overline{\rho_{12}}$ is also real (since $\tau_4=0$), we obtain that
$$
\mathrm{Re}\left(|a_1|^2+|a_2|^2+|c_1|^2+|c_2|^2-2\overline{a_1}a_2-2\overline{c_1}c_2\right)=0,
$$
which happens if and only if $a_1=a_2$ and $c_1=c_2.$ Furthermore, $A=aI_2$ and $C=cI_2,$ where $a_1=a_2=a\in\mathbb{C}$ and $c_1=c_2=c\in\mathbb{C}$ and thus $H$ does not contribute with the recurrence of the walk.
\end{proof}

\begin{teo}[Recurrence criteria for coins of dimension 2]\label{2EiCriteria}
Consider a CTOQW induced by a coin $(C,A)_H$ of dimension 2, and let $$\ket{u_1}=\begin{bmatrix}1\\0\end{bmatrix},\;\ket{u_2}=\begin{bmatrix}0\\1\end{bmatrix}.$$

(1) If $\mathbb{L}$ has a unique stationary state $\rho_{inv}$, then
\begin{itemize}
  \item $\mathrm{Tr}(A\rho_{inv}A^*)\neq\mathrm{Tr}(C\rho_{inv}C^*)\Rightarrow$ the walk is transient;
  \item $\mathrm{Tr}(A\rho_{inv}A^*)=\mathrm{Tr}(C\rho_{inv}C^*)\Rightarrow$ the walk is recurrent.
\end{itemize}

\medskip

(2) If $\mathbb{L}$ has more than one stationary state, then $C$ and $A$ share two orthogonal eigenvectors, and $H$ does not contribute with the recurrence of the walk.  In this case, we can represent $C,A,H$ by
\begin{equation*}
\begin{split}
C=& \begin{bmatrix}c_1 & 0 \\0 & c_2\end{bmatrix},\;
A=\begin{bmatrix}a_1 & 0 \\0 & a_2\end{bmatrix},\;
\quad a_j, \;c_j\in \mathbb{C},\quad j=1,2, \\
H=& \begin{bmatrix}h_1 & h_2 \\\overline{h_2} & h_3\end{bmatrix},\;h_2\in \mathbb{C},\quad h_1,\;h_3\in\mathbb{R}.
\end{split}
\end{equation*}
In this case, the two common orthonormal eigenvectors of $C,A$ are $\ket{u_1},\ket{u_2}.$

(2.1) If $h_2= 0,$ then
\begin{description}
  \item[a] $|a_i|\neq|c_i|$ for $i=1,2\;\Rightarrow$  $(C,A)_H$ is transient;
  \item[b] $|a_i|=|c_i|$  for $i=1,2\;\Rightarrow$ $(C,A)_H$ is recurrent;
  \item[c] $|a_i|=|c_i|,\;|a_j|\neq|c_j|,\;i\neq j,\;i,j\in\{1,2\}\Rightarrow$ $(C,A)_H$ is $\ket{u_j}\bra{u_j}$-transient and $\rho$-recurrent for all densities $\rho\neq\ket{u_j}\bra{u_j}$.
\end{description}

(2.2) If $h_2\neq 0,$ then $a_1=a_2=a,\; c_1=c_2=c$ for some $a,c\in \mathbb{C}^2.$ In this case,
\begin{itemize}
  \item $a=c\Rightarrow$ the walk is recurrent;
  \item $a\neq c\Rightarrow$ the walk is transient.
\end{itemize}
\begin{proof}
Item (1) is a particular case of Theorem \ref{corR}.

Now we suppose that $\mathbb{L}$ has more than one stationary state. By Proposition \ref{prop2inv}, $C$ and $A$ are diagonal with respect to some orthonormal basis of $\mathbb{C}^2,$ thus they are normal and share two orthonormal eigenvectors, namely $\ket{u_1},\ket{u_2}\in\mathbb{C}^2.$ By the spectral theorem, we can assume that
$$
C= \begin{bmatrix}c_1 & 0 \\0 & c_2\end{bmatrix},\;
A=\begin{bmatrix}a_1 & 0 \\0 & a_2\end{bmatrix},\;
\quad a_j, \;c_j\in \mathbb{C},\quad j=1,2,
$$
in the basis $\beta:=\{\ket{u_1},\ket{u_2}\}$ of $\mathbb{C}^2.$

Letting
$$
H= \begin{bmatrix}h_1 & h_2 \\\overline{h_2} & h_3\end{bmatrix},\; \quad h_1,\;h_3\in\mathbb{R},\quad h_2\in \mathbb{C},
$$
in $\beta$, we first suppose that $h_2=0$ and either $a_1\neq a_2$ or $c_1\neq c_2.$

Consider the set
$$
\hat{\mathcal{D}}:=\left\{\rho\in\mathcal{S}(\mathcal{H}):\rho=\sum_{i\in V}\rho(i)\otimes\ket{i}\bra{i},\rho(i)\in\mathbb{M}_2(\mathbb{C})\mbox{ is diagonal for each i in basis }\beta\right\},
$$
which is a subset of $\mathcal{D}.$ Note that $\hat{\mathcal{D}}$ represents the set of all block-diagonal densities such that each $\rho(i)\in\mathbb{M}_2(\mathbb{C})$ can be represented by a diagonal matrix in basis $\beta.$

If $\rho\in\hat{\mathcal{D}},$ then is straightforward that $\mathcal{L}(\rho\otimes\ket{i}\bra{i})\in\hat{\mathcal{D}},$ since $C,A,H$ are all diagonal. Therefore $\mathcal{L}^2(\rho\otimes\ket{i}\bra{i}),\;\mathcal{L}^3(\rho\otimes\ket{i}\bra{i}),\;\mathcal{L}^4(\rho\otimes\ket{i}\bra{i}),\ldots\in\hat{\mathcal{D}},$ and so on we see that
$$
e^{t\mathcal{L}}(\rho\otimes\ket{i}\bra{i})=\sum_{n=0}^{\infty}\frac{t^n}{n!}\mathcal{L}^n(\rho\otimes\ket{i}\bra{i})\in\hat{\mathcal{D}}.
$$

We have
$$
C=\begin{bmatrix}
    c_1 & 0 \\
    0 & c_2
  \end{bmatrix},\qquad
A=\begin{bmatrix}
    a_1 & 0 \\
    0 & a_2
  \end{bmatrix},\;\;a_j,\;c_j\in \mathbb{C},\qquad
\ket{u_1}=\begin{bmatrix}1 \\0\end{bmatrix}\qquad
\ket{u_2}=\begin{bmatrix}0 \\1\end{bmatrix},\qquad
$$
thus
$$A\ket{u_j}=a_j,\;C\ket{u_j}=c_j,\;k=1,2.$$

Consider the densities
$$\rho_1=\ket{u_1}\bra{u_1}=
\begin{bmatrix}1 & 0 \\0 & 0\end{bmatrix},
\qquad
\rho_2=\ket{u_2}\bra{u_2}=\begin{bmatrix}0 & 0 \\0 & 1\end{bmatrix}.$$
As discussed above, $e^{t\mathcal{L}}(\rho_1\otimes\ket{i}\bra{i})\in\hat{\mathcal{D}},$ showing that $\rho_t(i),$ as defined in Equation \eqref{rho_t(i)}, is of the form
$$
\rho_t(i)=\begin{bmatrix}
            \tau_t & 0 \\
            0 & 0
          \end{bmatrix},\qquad 0<\tau_t<1,
$$
leading us to the probability
$$
p_{ii;\rho_1}(t)=\mathrm{Tr}(\rho_t(i)\otimes\ket{i}\bra{i})=\tau_t.
$$

So we can note that if we start the walk at vertex $i$ with initial density operator $\rho_1,$ the values $c_2$ and $a_2$ do not contribute with the transition probabilities. Moreover, the operator $H$ does not contribute with the transitions, since
$$
G_i\rho_1+\rho_1 G_i^*=-iH_i\rho_1-\frac{1}{2}\sum_{j\in V}R_i^{j*}R_i^j\rho_1+i\rho_1 H_i-\frac{1}{2}\rho_1\sum_{j\in V}R_i^{j*}R_i^j
=-\sum_{j\in V}R_i^{j*}R_i^j\rho_1,
$$
where $H$ is hermitian and the commutative property holds because all the involved matrices $\rho_1$ and $R_i^{j*}R_i^j$ are diagonal and hermitian. Now it is trivial that the CTOQW is $\rho_1$-recurrent if and only if the continuous-time Markov chain with $Q$-matrix\footnote{Some authors call a $Q$-matrix as the infinitesimal operator matrix of the continuous-time Markov chain \cite{manuel}.}
$$
Q=
\begin{bmatrix}
  \ddots & \ddots & \ddots & & &  & 0 \\
   & |a_1|^2 & -|a_1|^2-|c_1|^2 & |c_1|^2 &  &&  \\
&& |a_1|^2  & -|a_1|^2-|c_1|^2 & |c_1|^2 &  &  \\
  &&& |a_1|^2  & -|a_1|^2-|c_1|^2 & |c_1|^2 &   \\
 0 &&& & \ddots & \ddots&\ddots
\end{bmatrix}
$$
is recurrent, which is recurrent only for $|a_1|=|c_1|$ \cite[Example 3.66]{manuel}. See \cite{ander,norris} for more on $Q$-matrices. An analogous result can be obtained for $\rho_2$ and the values $|a_2|,\;|c_2|.$

In short,
$$(C,A)_H \mbox{ is }\rho_j\mbox{-recurrent if and only if }|a_j|=|c_j|.$$

Let us assume the walk is neither recurrent nor transient, then by items 3 and 4 of Proposition \ref{discussrec}, the walk is
$$
\ket{v_1}\bra{v_1}\mbox{-recurrent and}\ket{v_2}\bra{v_2}\mbox{-transient}
$$
for some $\ket{v_1},\ket{v_2}\in\mathbb{C}^2.$ This implies that $(C,A)_H$ is $\ket{v_2}\bra{v_2}$-transient and $\sigma$-recurrent for all $\sigma\neq\ket{v_2}\bra{v_2},$ otherwise
$$
\mathcal{Y}=\{\rho:(C,A)_H\mbox{ is }\rho\mbox{-transient}\}\subset \mathbb{C}^2
$$
would be an invariant subspace of dimension 2.

Since the CTOQW is homogeneous, $\ket{v_2}$ must be a common eigenvector of $C$ and $A.$ Indeed, if the initial density operator is $\rho=\ket{v_2}\bra{v_2},$ we must have $\rho=\rho_{T_n}$ for all $n\in\mathbb{N},$ otherwise the walk would be $\rho_{T_n}$-recurrent, since we could refresh the walk after the $n$-th jump and the density operator at this stage would be $\rho_{T_n},$ which is a contradiction (see Remark \ref{RemInvSub}).

We conclude that $\ket{v_2}$ is a multiple of $\ket{u_j}$ for some $j\in\{1,2\},$ and consequently the walk is neither recurrent nor transient when or $|a_1|=|c_1|$ and $|a_2|\neq|c_2|$ or $|a_2|=|c_2|$ and $|a_1|\neq|c_1|$. Finally, in this case, Proposition \ref{discussrec} assures that $(C,A)_H$ is transient with respect to a unique initial density operator and recurrent with respect to all the other density operators according to the modulus of $|c_j|$ and $|a_j|,\;j=1,2.$

Now we let $h_2\neq 0.$ By Lemma \ref{lemmaH}, $H$ does not contribute with the recurrence of the CTOQW, $a=a_1=a_2$ and $c=c_1=c_2$, thus the arguments above show that the walk is recurrent if and only if $|a|=|c|.$
\end{proof}
\end{teo}

The last result of this work is a consequence of the Theorem above. It shows that the recurrence of a CTOQW induced by a coin $(C,A)_H$ of dimension 2 with $C$ and $A$ being diagonal in a common basis depends only on the modulus of the elements on the diagonal (also the modulus of the eigenvalues) of $C$ and $A$, once we know if $\rho_{inv}$ is unique.

\begin{prop}\label{LastProp}
Consider a CTOQW induced by a coin $(C,A)_H$ such that there exists a basis in which
$$
C=\begin{bmatrix}
    c_1 & 0 \\
    0 & c_2
  \end{bmatrix},\qquad
A=\begin{bmatrix}
    a_1 & 0 \\
    0 & a_2
  \end{bmatrix},\;\;a_j,\;c_j\in \mathbb{C},\; j=1,2.
$$
The Hamiltonian $H$ determines if $\rho_{inv}$ is unique, and after that, it does not contribute with the recurrence of the walk and we are in one of the following situations:
\begin{description}
  \item[(i)]If $\mathbb{L}$ has only one stationary state $\rho_{inv}$, then $\rho_{inv}=I_2$ and

  \medskip

  \begin{itemize}
    \item $|a_1|^2+|a_2|^2=|c_1|^2+|c_2|^2\Rightarrow\;(C,A)_H$ is recurrent;

    \medskip

    \item$|a_1|^2+|a_2|^2\neq|c_1|^2+|c_2|^2\Rightarrow\;(C,A)_H$ is transient.
  \end{itemize}

  \medskip

  \item[(ii)] If $\mathbb{L}$ has more than one stationary state, then we are in situation (2) of Theorem \ref{2EiCriteria}.
\end{description}
\begin{proof}
We just have to prove the case (i). So, let us suppose that $\rho_{inv}$ is unique, then
$$ \rho_{11}=\frac{1}{2}\quad \mbox{ and }\quad \rho_{12}=0 $$
is a solution for Equation \eqref{taus}, thus
$$\rho_{inv}=\frac{1}{2}\begin{bmatrix}
1 & 0 \\
0 & 1
\end{bmatrix}$$
is the only stationary state of $\mathbb{L}.$
A simple calculation gives
$$
\mathrm{Tr}\left(A\rho_{inv}A^*\right)-\mathrm{Tr}\left(C\rho_{inv}C^*\right)=\left(|a_1|^2+|a_2|^2\right)-\left(|c_1|^2+|c_2|^2\right),
$$
which is null if and only if $|a_1|^2+|a_2|^2=|c_1|^2+|c_2|^2.$ The proof is finished by an application of Theorem \ref{2EiCriteria}, item (1).
\end{proof}
\end{prop}

\section{Examples}
In this section, we present examples of CTOQWs induced by different coins and describe how to verify their recurrence. The stationary states were computed with Equation \eqref{LforCoin}, and some of the most exhaustive calculations were obtained with Maple 15 software.

\begin{ex}
Let us consider a CTOQW induced by the coin $(C,A)_H,$
$$
C=\begin{bmatrix}
    \sqrt{2} & 0 \\
    0 & \sqrt{11}
  \end{bmatrix},\quad
A=\begin{bmatrix}
    -\sqrt{5} & 0 \\
    0 & a
  \end{bmatrix},\quad
H=\begin{bmatrix}
  1 & 1-2i \\
  1+2i & 1
\end{bmatrix},
\quad a\in \mathbb{C}.
$$
Since $H$ is not diagonal, $\mathbb{L}$ has a unique stationary state $\rho_{inv}=I_2.$ Indeed, if $\mathbb{L}$ had at least two stationary states, we would be in situation (2.1) or (2.2) of Theorem \ref{2EiCriteria}, which is impossible. Therefore, we are in situation (i) of Proposition \ref{LastProp} (also situation (1) of Theorem \ref{2EiCriteria}), from where we conclude that
 \begin{itemize}
    \item $(C,A)_H$ is recurrent $\Leftrightarrow 5+|a|^2=13\Leftrightarrow |a|=2\sqrt{2};$

    \medskip

    \item $(C,A)_H$ is transient $\Leftrightarrow |a|\neq 2\sqrt{2}.$
  \end{itemize}
\qee
\end{ex}

\begin{ex}
Consider the CTOQW induced by the coin $(C,A)_H,$
$$
C=\frac{1}{2}\begin{bmatrix}
    1+c & i(-1+c) \\
    i(1-c) & 1+c
  \end{bmatrix},\quad
A=\frac{1}{2}\begin{bmatrix}
    2+a & i(2-a) \\
    i(-2+a) & 2+a
  \end{bmatrix},\;\; i=\sqrt{-1},\quad a,\;c\in \mathbb{C},
$$
and
$$
H= \begin{bmatrix}h_1 & h_2 \\\overline{h_2} & h_3\end{bmatrix},\;h_2\in \mathbb{C},\quad h_1,\;h_3\in\mathbb{R}.
$$

The matrices $C$ and $A$ share two orthogonal eigenvectors
$$\ket{u_1}=\frac{\sqrt{2}}{2}[-i,1]^t,\quad\ket{u_2}=\frac{\sqrt{2}}{2}[i,1]^t$$ with $C\ket{u_1}=\ket{u_1},\;C\ket{u_2}=c\ket{u_2},\; A\ket{u_1}=a\ket{u_1},\;A\ket{u_2}=2\ket{u_2}.$
By the Spectral Theorem, we have
$$
C=U\begin{bmatrix}
    1 & 0 \\
    0 & c
  \end{bmatrix}U^*,\quad
A=U\begin{bmatrix}
    a & 0 \\
    0 & 2
  \end{bmatrix}U^*,\quad
H=U\left( U^*\begin{bmatrix}h_1 & h_2 \\\overline{h_2} & h_3\end{bmatrix}U\right)U^*,
$$
where $U$ is the unitary matrix of $\mathbb{C}^2$ such that $U\ket{e_1}=\ket{u_1},\;U\ket{e_2}=\ket{u_2},$ and $\{\ket{e_1},\ket{e_2}\}$ is the canonical basis of $\mathbb{C}^2.$ In matrix terms,
$$
U=\frac{\sqrt{2}}{2}
\begin{bmatrix}
  -i & i \\
  1 & 1
\end{bmatrix}.
$$
Therefore, we have
$$
C=\begin{bmatrix}
    1 & 0 \\
    0 & c
  \end{bmatrix},\quad
A=\begin{bmatrix}
    a & 0 \\
    0 & 2
  \end{bmatrix},\quad
H=\frac{1}{2}\begin{bmatrix}h_1-2\mathrm{Re}(h_2)+h_3 & -i\left(h_1+2\mathrm{Im}(h_2)-h_3\right) \\ i\left(h_1-2\mathrm{Im}(h_2)-h_3\right) & h_1+2\mathrm{Re}(h_2)+h_3\end{bmatrix},
$$
in basis $\{\ket{u_1},\ket{u_2}\}.$ Firstly, let us suppose that
$$h_1+2\mathrm{Im}(h_2)-h_3=0,$$
then the antidiagonal of $H$ (in the basis $\{\ket{u_1},\ket{u_2}\}$) is null and $\mathbb{L}$ has more than one stationary state. The values $a_j,\;c_j,\;j=1,2,$ appearing at item (2.1) of Theorem \ref{2EiCriteria} are
$$
c_1=1,\quad c_2=c,\quad
a_1=a,\quad a_2=2.
$$

We have
\begin{description}
  \item[a] If $|a|\neq 1$ and $|c|\neq 2,$  then the walk is transient;
  \item[b] If $|a|= 1$ and $|c|= 2,$ then the walk is recurrent;
  \item[c] If $|a|\neq 1$ and $|c|=2,$ then $(C,A)_H$ is $\ket{u_1}\bra{u_1}$-transient and $\rho$-recurrent for all densities $\rho\neq\ket{u_1}\bra{u_1}$;
  \item[d] If $|a|= 1$ and $|c|\neq 2,$ then $(C,A)_H$ is $\ket{u_2}\bra{u_2}$-transient and $\rho$-recurrent for all densities $\rho\neq\ket{u_2}\bra{u_2}$.
  \end{description}

On the other hand, if $h_1+2\mathrm{Im}(h_2)-h_3\neq 0,$ then $\mathbb{L}$ has only one stationary state and by Proposition \ref{LastProp}, item (i), $(C,A)_H$ is recurrent for
$$|c|^2-|a|^2=3$$
and transient otherwise.
\qee
\end{ex}

\begin{ex}\label{Ex2}
Let us consider the CTOQW induced by the coin $(C_y,A_y)_{H_h},$
$$
C_y=\begin{bmatrix}
    -1 & 1 \\
    2y & 1
  \end{bmatrix},\quad
A_y=\begin{bmatrix}
    1 & 1 \\
    y & 2
  \end{bmatrix},\quad
H_h=\begin{bmatrix}
    0 & i h \\
    -i h & 0
  \end{bmatrix},\quad
y,h\in \mathbb{R}.
$$

For simplicity, we assume that $|y|\neq 1$ and $25y^4+8h^2-30y^3-12yh+37y^2+12h-36y+14\neq 0.$ The reason for this hypothesis will be cleared later.

The invariant state of $\mathbb{L}$ is $\rho_{inv}=\begin{bmatrix}a & b \\b & 1-a\end{bmatrix},$ where
\begin{equation*}
\begin{split}
a=& \frac{2(7+6h-18y-8yh+13y^2+2h^2)}{25y^4+8h^2-30y^3-12yh+37y^2+12h-36y+14},\\
b=& \frac{2(-10y^3+15y^2+5y^2h-6y-2h)}{25y^4+8h^2-30y^3-12yh+37y^2+12h-36y+14}.
\end{split}
\end{equation*}

A calculation gives
\begin{equation*}
\begin{split}
m=& \mathrm{Tr}(A_y\rho_{inv}A_y^*)-\mathrm{Tr}(C_y\rho_{inv}C_y) \\
     =& \frac{12yh-16h+48y^3h-12y^2h^2+111y^2+12h^2-3y^4-62y^3+4y^2h}{25y^4+8h^2-30y^3-12yh+37y^2+12h-36y+14}.
\end{split}
\end{equation*}

Firstly, we assume that $y=0.$ In this case, $C_0$ and $A_0$ share only one eigenvector, namely $\ket{u}=[1,0]^t,$ thus by Theorem \ref{2EiCriteria}, the walk is recurrent (transient) if and only if $m=0 \;(m\neq 0).$ The value $m$ is explicitly
$$
m=\frac{2h(3h-4)}{4h^2+6h+7}\begin{cases}
                              =0, & \mbox{if } h\in\{0,4/3\} \\
                              \neq 0, & \mbox{otherwise}.
                            \end{cases}
$$
Therefore $(C_0,A_0)_{H_h}$ is recurrent if and only if $h=0$ or $h=4/3.$ Otherwise, the walk is transient.

Secondly, let us assume that $y\neq 0.$ In this case, $C_y$ and $A_y$ share no eigenvector, thus by Theorem \ref{2EiCriteria}, the walk is recurrent if and only if $m=0,$ which happens when
$$
h=\frac{y^2-4+3y+12y^3\pm\sqrt{415y^4-332y^2-48y^3-162y^5+16+120y+135y^6}}{6(y^2-1)}.
$$
Otherwise, the walk is transient.

In particular, taking $y=1/2$ we conclude that the coin $(C_{\frac{1}{2}},A_{\frac{1}{2}})_{H_h}$ is recurrent for
$$h=\frac{2\pm\sqrt{71}}{12}$$
and otherwise, it is transient.
\qee
\end{ex}

\begin{ex}
Let us finish this section with an example of CTOQW induced by a coin $(C,A)_H$ of dimension $3.$ We consider
$$
C=\begin{bmatrix}
  c & 0 & 0 \\
  1 & 0 & 0 \\
  0 & 0 & 1
\end{bmatrix},\;c\in\mathbb{R},\quad
A=\begin{bmatrix}
  1 & 1 & 0 \\
  0 & 0 & 1 \\
  0 & 0 & 1
\end{bmatrix},\quad
H_i=
\begin{bmatrix}
  1 & 2 & 0 \\
  2 & 0 & 0 \\
  0 & 0 & 0
\end{bmatrix},\quad i\in \mathbb{R}.
$$

If $c=0,$ then
$$
\rho_{inv}=\frac{1}{53}
\begin{bmatrix}
  21 & -19-2i & 0 \\
  -19+2i & 32 & 0 \\
  0 & 0 & 0
\end{bmatrix}
$$
and $m=\mathrm{Tr}(A\rho_{inv}A^*)-\mathrm{Tr}(C\rho_{inv}C)=-6/53.$

If $c=1,$ then
$$
\rho_{inv}=\frac{1}{2}
\begin{bmatrix}
  1 & 0 & 0 \\
  0 & 1 & 0 \\
  0 & 0 & 0
\end{bmatrix}
$$
and $m=0$.

By Theorem \ref{corR}, we conclude that
\begin{equation*}
\begin{split}
 \bullet\;\;  c=0 \Rightarrow&\; \mbox{the walk is transient}; \\
 \bullet\;\;   c=1 \Rightarrow&\; \mbox{the walk is recurrent}.
\end{split}
\end{equation*}

For completeness, it is possible to show that this CTOQW is recurrent if and only if $c=1.$
\qee
\end{ex}

\textbf{Acknowledgements:} The author is grateful to an anonymous referee for many useful suggestions that led to a marked improvement of the paper and acknowledges financial support by CAPES (Coordena\c{c}\~ao de Aperfei\c{c}oamento de Pessoal de N\'ivel Superior) during the period 2018-2022.

\bigskip

\section*{Appendices}

\textbf{Proof of Proposition} \ref{lard}.
Let $i,j\in V$ and $\alpha,\beta\geq 0.$ For simplicity, we will denote $P_k=I\otimes\ket{k}\bra{k}$ for $k\in V,$ then we use the semigroup property to obtain
\begin{equation*}
\begin{split}
p_{ji;\rho}(\alpha+\beta) =&\mbox{Tr}\left[e^{(\alpha+\beta)\mathcal{L}}(\rho\otimes\ket{i}\bra{i})P_j\right]=
\mbox{Tr}\left[e^{\beta\mathcal{L}}(\rho\otimes\ket{i}\bra{i})e^{\alpha\mathcal{L}^*}(P_j)\right]  \\
=&\sum_k\mbox{Tr}\left[e^{\beta\mathcal{L}}(\rho\otimes\ket{i}\bra{i})P_k\;e^{\alpha\mathcal{L}^*}(P_j)\right]  \\
=&\sum_k\mbox{Tr}\left[e^{\alpha\mathcal{L}}\left(e^{\beta\mathcal{L}}(\rho\otimes\ket{i}\bra{i})P_k\right)P_j\right]\\
=&\sum_k\mbox{Tr}\left[e^{\alpha\mathcal{L}}\left(\frac{e^{\beta\mathcal{L}}(\rho\otimes\ket{i}\bra{i})P_k}{\mbox{Tr}(e^{\beta\mathcal{L}}(\rho\otimes\ket{i}\bra{i})P_k)}\right)P_j\right]
\mbox{Tr}(e^{\beta\mathcal{L}}(\rho\otimes\ket{i}\bra{i})P_k)\\
=&\sum_{k}p_{jk;\rho'_{ki}(\beta)}(\alpha)p_{ki;\rho}(\beta),
\end{split}
\end{equation*}
where
$$
\rho'_{ki}(\beta)=\frac{e^{\beta\mathcal{L}}(\rho\otimes\ket{i}\bra{i})P_k}{\mbox{Tr}(e^{\beta\mathcal{L}}(\rho\otimes\ket{i}\bra{i})P_k)}
$$
is a density operator.
$\square$

\smallskip
\textbf{Proof of Proposition} \ref{jointlycontinuous}.
Define the function $g:W_i\rightarrow [0,1]$ by $g(\rho,s)=p_{ji;\rho}(s).$ Since $e^{t\mathcal{L}}$ is uniformly continuous,  $g$ is continuous on $[0,+\infty)$ for a fixed $\rho\in S_{\mathfrak{h}_i},$ thus, given $\varepsilon>0,$ there is an $\alpha>0$ such
that $|t-s|<\alpha$ implies $|g(\rho,t)-g(\rho,s)|<\varepsilon/2.$

Let $\beta:=\min(\alpha,\varepsilon/2).$ If $|t-s|<\beta$ and $\|\rho-\rho'\|_1<\beta,$ where $\|\cdot\|_1$ is the trace norm in $\mathfrak{h}_i,$ we have
\begin{equation*}
\begin{split}
|g(\rho,s)-g(\rho',s)|=&\left|\mbox{Tr}\left[e^{s\mathcal{L}}((\rho-\rho')\otimes\ket{i}\bra{i})P_j\right]\right| \\
=&\left|\mbox{Tr}\left[((\rho-\rho')\otimes\ket{i}\bra{i})e^{s\mathcal{L}^*}(P_j)\right]\right|\\
\leq&\left\|(\rho-\rho')\otimes\ket{i}\bra{i}\right\|_1\;\left\|e^{s\mathcal{L}^*}(P_j)\right\|\\
<&\beta,
\end{split}
\end{equation*}
thus we obtain
$$
|g(\rho,t)-g(\rho',s)| \leq |g(\rho,t)-g(\rho,s)|+|g(\rho,s)-g(\rho',s)|<\varepsilon.
$$
This concludes the proof.
$\square$

\smallskip
\textbf{Proof of Proposition} \ref{cont and min}.
(1)By contradiction, suppose that there exists $k>0$ with $p_{ii;\rho}(k)=0.$ Since $p_{ii;\rho}(t)$ is jointly continuous on $(t,\rho)\in\left([0,\infty)\times\mathcal{S}(\mathfrak{h}_i)\right)$ and
$p_{ii;\rho}(0)=1,$ we can assume that $k=\min\{s>0:p_{ii;\rho}(s)=0\}.$ Moreover, there exists $\varepsilon>0$ such that, for $t<k,$
\begin{equation*}\label{limcon}
k-t<\varepsilon\mbox{ and }\|\rho-\tilde{\rho}\|<\varepsilon\Rightarrow p_{ii;\tilde{\rho}}(t)>0.
\end{equation*}
Now, note that
$$
\rho'_{ii}(k/n)=\frac{e^{\frac{k}{n}\mathcal{L}}(\rho\otimes\ket{i}\bra{i})P_i}{\mbox{Tr}(e^{\frac{k}{n}\mathcal{L}}(\rho\otimes\ket{i}\bra{i})P_i)}
\stackrel{n\rightarrow\infty}{\longrightarrow}
\frac{(\rho\otimes\ket{i}\bra{i})P_i}{\mbox{Tr}((\rho\otimes\ket{i}\bra{i})P_i)}=\rho\otimes\ket{i}\bra{i}.
$$

Now, take $n>0$ such that $\frac{k}{n}<\varepsilon$ and $\|\rho\otimes\ket{i}\bra{i}-\rho'_{ii}(k/n)\|<\varepsilon,$ then $p_{ii;\rho'_{ii}(k/n)}((kn-k)/n)>0,$
thus
$$p_{ii;\rho}(k)\geq p_{ii;\rho'_{ii}(k/n)}((kn-k)/n)p_{ii;\rho}(k/n)>0
$$
holds by Proposition \ref{chapquantum}, which gives a contradiction.

For item (2), let $x\geq 0,$ then item (1) gives
$$p_{ij;\rho}(t+x)\geq p_{ii;\rho'_{ij}(t)}(x)p_{ij;\rho}(t)>0.$$

Suppose $p_{ii;\rho}(t)=1$ for some $t>0.$ If we had $p_{ji;\rho}(s)>0$ for some $j\neq i$ and $s\in[0,t],$ then
$$0=\sum_{k\neq i}p_{ki;\rho}(t)\geq p_{ji;\rho}(t-s+s)\geq p_{ii;\rho'_{ji}(s)}(t-s)p_{ji;\rho}(s)>0,$$
which is a contradiction. This shows item (3).

To prove item (4), note that for fixed $0\leq t_0<\delta,$ $W(i,\delta):=S({\mathfrak{h}_i})\times[t_0,\delta]$ is a compact set in $W_i=S(\mathfrak{h}_i)\times[0,+\infty)$. Hence, by the jointly continuity, $N_{ji}$ is attained on $(0,1].$
$\square$

\smallskip
\textbf{Proof of Theorem} \ref{equiv}.
If $p_{ji;\rho}(t)=0$ for all $t,$ then the result is obvious. Thus suppose $p_{ji;\rho}(t)>0,$ for some $t\geq 0.$ Item (2) of Proposition
\ref{cont and min} assures the existence of $M_\delta\in\mathbb{N}$ such that $p_{ji;\rho}(n\delta)>0,\;\forall n\geq M_\delta.$

By the Mean Value Theorem for Integrals, we have
$$\int_{0}^{\infty}p_{ji;\rho}(t)dt=\sum_{n=0}^{\infty}\int_{n\delta}^{(n+1)\delta}p_{ji;\rho}(t)dt=\sum_{n=0}^{\infty}\delta
p_{ji;\rho}(n\delta+s_n),$$
 where $(s_n)_{n=0}^\infty$ is a sequence in $[0,\delta].$

By Proposition \ref{lard},
\begin{equation}\label{des11}
p_{ji;\rho}(n\delta+s_n)\geq p_{ji;\rho}(n\delta)p_{jj;\rho'_{ji}(n\delta)}(s_n),\quad\forall n\geq M_\delta,
\end{equation}
 and
\begin{equation}\label{des22}
p_{ji;\rho}(n\delta+\delta)=p_{ji;\rho}(n\delta+s_n+\delta-s_n)\geq p_{jj;\rho'_{ji}(n\delta+s_n)}(\delta-s_n)p_{ji;\rho}(n\delta+s_n), \quad\forall n\geq M_\delta,
\end{equation}
so that for any fixed $\rho$,
\begin{equation}\label{intser}
\begin{split}
\int_{0}^{\infty}p_{ji;\rho}(t)dt=&\; \delta\sum_{n=0}^{\infty} p_{ji;\rho}(n\delta+s_n)dt\geq \delta\sum_{n=M_\delta}^{\infty} p_{ji;\rho}(n\delta+s_n)dt \\
\stackrel{\eqref{des11}}{\geq}&\; \delta\sum_{n=M_\delta}^{\infty} p_{ji;\rho}(n\delta)p_{jj;\rho'_{ji}(n\delta)}(s_n)\geq  \delta N_{jj}\sum_{n=M_\delta}^{\infty} p_{ji;\rho}(n\delta)
\end{split}
\end{equation}
and
\begin{equation}\label{serint}
\begin{split}
\sum_{n=0}^{\infty} p_{ji;\rho}(n\delta+\delta)\stackrel{\eqref{des22}}{\geq}&\; \sum_{n=0}^{\infty}p_{jj;\rho'_{ji}(n\delta+s_n)}(\delta-s_n)p_{ji;\rho}(n\delta+s_n) \\
\geq&\; \sum_{n=M_\delta}^{\infty}p_{jj;\rho'_{ji}(n\delta+s_n)}(\delta-s_n)p_{ji;\rho}(n\delta+s_n) \\
\geq&\; N_{jj}\sum_{n=M_\delta}^{\infty}p_{ji;\rho}(n\delta+s_n) \\
=&\;  \frac{N_{jj}}{\delta}\sum_{n=M_\delta}^{\infty}\delta p_{ji;\rho}(n\delta+s_n) \\
=&\;\frac{N_{jj}}{\delta}\sum_{n=0}^{\infty}\delta p_{ji;\rho}(n\delta+s_n)-\frac{N_{jj}}{\delta}\sum_{n=0}^{M_\delta-1}\delta p_{ji;\rho}(n\delta+s_n) \\
=&\;\frac{N_{jj}}{\delta}\int_{0}^{\infty}p_{ji;\rho}(t)dt-\frac{N_{jj}}{\delta}\sum_{n=0}^{M_\delta-1}\delta p_{ji;\rho}(n\delta+s_n).
\end{split}
\end{equation}

Whence, for a state $\rho$, the divergence of the series in \eqref{intser} implies the divergence of the integral on the left. Also, if
we suppose the integral on \eqref{serint} diverges, then the series on the left diverges since the series on the right-hand is finite.
$\square$

\smallskip
\textbf{Proof of Proposition} \ref{discussrec}.
 $1.$ Since $\rho$ is faithful, there exists $\alpha>0$ such that $\rho\geq\alpha\tilde{\rho},$ thus
 \begin{equation*}
\begin{split}
\int_{0}^{\infty}p_{ii;\rho}(t)dt=& \int_{0}^{\infty}\mbox{Tr}\left[e^{t\mathcal{L}}(\rho\otimes\ket{i}\bra{i})P_i\right]dt \\
\geq& \alpha\int_{0}^{\infty}\mbox{Tr}\left[e^{t\mathcal{L}}(\tilde{\rho}\otimes\ket{i}\bra{i})P_i\right]dt \\
=&\alpha\int_{0}^{\infty}p_{ii;\tilde{\rho}}(t)dt\\
=&\infty.
\end{split}
 \end{equation*}

$2.$ Suppose $\rho'_{ii}(\delta)$ is faithful for some $\delta\geq 0.$ The item $1$ gives that $i$ is $\rho'_{ii}(\delta)$-recurrent, thus
$$
\int_{0}^{\infty}p_{ii;\rho}(t)dt\geq\int_{0}^{\infty}p_{ii;\rho}(t+\delta)dt\geq\int_{0}^{\infty}p_{ii;\rho}(\delta)p_{ii;\rho'_{ii}(\delta)}(t)dt=p_{ii;\rho}(\delta)\int_{0}^{\infty}p_{ii;\rho'_{ii}(\delta)}(t)dt=\infty.
$$

$3.$ Let $\rho\in\mathcal{S}(\mathfrak{h}_i)$ be non-faithful. By the Spectral Theorem, $\rho$ can be written as
\begin{equation}\label{specde}
\rho=\sum_{x=1}^{n}\lambda_x\ket{x}\bra{x},
\end{equation}
where the vectors $\ket{x},\;x=1,\ldots,n,$ are the eigenvectors of $\rho$ with eigenvalues $\lambda_x,\;x=1,\ldots,n.$ Since $\rho$ is non-faithful, there is at least one null eigenvalue
and the remainder eigenvalues are positive summing $1.$ Thus, \eqref{specde} can be rewritten as
\begin{equation*}\label{spec2}
\rho=\sum_{x\in S}\lambda_x\ket{x}\bra{x}, S\varsubsetneq \{1,\ldots,n\}.
\end{equation*}
Take a sequence of positive numbers $(\alpha_r)_{r\in R},$ where $R:=\{1,\ldots,n\}/S\neq \emptyset,$ whose sum is $1.$

Defining the density operator
$$\rho_X=\sum_{x\in S}\frac{\lambda_x}{2}\ket{x}\bra{x}+\sum_{x\in
R}\frac{\alpha_x}{2}\ket{x}\bra{x}=\sum_{x=1}^{n}\frac{\tilde{\alpha}_x}{2}\ket{x}\bra{x},\;\tilde{\alpha}_x=\begin{cases}
                                                                                                             \lambda_x, & \mbox{if } x\in S \\
                                                                                                             \alpha_x, & \mbox{if } x\in R
                                                                                                           \end{cases},$$
we get by item (1) that $i$ is $\rho_X$-recurrent, since $\rho_X$ is faithful.

Now, define
$$\rho_Y=\sum_{x\in R}\alpha_x\ket{x}\bra{x},$$
which is a non-faithful density operator and then we get $2\rho_X=\rho+\rho_Y.$ This leads us to
\begin{eqnarray}
\nonumber  \int_{0}^{\infty}p_{ii;\rho}(t)dt+\int_{0}^{\infty}p_{ii;\rho_Y}(t)dt &=& \int_{0}^{\infty}\left(p_{ii;\rho}(t)+p_{ii;\rho_Y}(t)\right)dt\\
\nonumber   &=& \int_{0}^{\infty}\left(\mbox{Tr}\left[e^{t\mathcal{L}}(\rho\otimes\ket{i}\bra{i})P_i\right]\right) +\mbox{Tr}\left[e^{t\mathcal{L}}(\rho_Y\otimes\ket{i}\bra{i})P_i\right])dt \\
\nonumber   &=& \int_{0}^{\infty}\mbox{Tr}\left[e^{t\mathcal{L}}((\rho+\rho_Y)\otimes\ket{i}\bra{i})P_i\right]\\
\nonumber   &=& 2\int_{0}^{\infty}\mbox{Tr}\left[e^{t\mathcal{L}}(\rho_X\otimes\ket{i}\bra{i})P_i\right]\\
\nonumber   &=& 2\int_{0}^{\infty}p_{ii;\rho_X}(t)dt.
\end{eqnarray}

The integral on the right-hand diverges once $i$ is $\rho_X$-recurrent. This implies that at least one of the integrals on the left-hand side
diverges. Therefore, $i$ is $\rho$-recurrent or $\rho_Y$-recurrent.

$4.$ By item (3), vertex $i$ is $\rho$-recurrent with respect to some non-faithful $\rho.$ By the spectral theorem, $\rho=\sum_j\lambda_j\ket{j}\bra{j},$ where each $\lambda_j$ is an eigenvalues of $\rho$ with corresponding eigenvector $\ket{j},$ thereby we have
$$
\infty=\int_{0}^{\infty}p_{ii;\rho}(t)dt=\sum_j\lambda_j\int_{0}^{\infty}p_{ii;\ket{j}\bra{j}}(t)dt,
$$
where the members of the sum are all positives, thus we have
$$\int_{0}^{\infty}p_{ii;\ket{j}\bra{j}}(t)dt=\infty\quad \mbox{ for some }\quad j.$$

$5.$ If $i$ is $\rho$-transient, then we can assume that $\rho=\ket{u}\bra{u}$ for some $\ket{u}\in\mathbb{C}^2$ by item 4. Now let $\ket{v}$ be any unitary vector of $\mathbb{C}^2$ linearly independent of $\ket{u},$ then $\rho=\frac{1}{2}\ket{u}\bra{u}+\frac{1}{2}\ket{v}\bra{v}$ is a faithful density operator and thus $i$ is $\rho$-recurrent, implying that $i$ is $\ket{v}\bra{v}$-recurrent. Since $\{\ket{u},\ket{v}\}$ forms a basis of $\mathbb{C}^2,$ $i$ must be recurrent with respect to every density different from $\rho.$
$\square$

\end{document}